\documentclass[12pt]{natureprintstyle}

\usepackage{cite}
\usepackage{times}
\usepackage{fullpage}
\usepackage{ifthen}
\usepackage{graphicx}
\usepackage[para]{footmisc}
\usepackage{amssymb}
\usepackage{amsmath}
\usepackage{color}
\usepackage{xcolor}
\usepackage{soul}
\usepackage{array}

\newcommand{\msolar} {$\rm{M_{\odot}}~$}
\newcommand{\msolarc} {$\rm{M_{\odot}}$}

\newcommand{\apj}{Astrophys. J.}   
\newcommand{\apjl}{Astrophys. J. Lett.}   
\newcommand{\apjs}{Astrophys. J. Suppl. Ser.}   
\newcommand{\aap}{Astron. Astrophys.}   
\newcommand{\joss}{J. Open Source Softw.}    
\newcommand{\lrr}{Living Rev. Relativ.}    
\newcommand{\mnras}{Mon. Not. R. Astron. Soc.}   
\newcommand{\nat}{Nature} 
\newcommand{\nastro}{Nat. Astron.} 
\newcommand{\pasj}{Publ. Astron. Soc. Jpn}   
\newcommand{\sci}{Science} 

\title{Growth of Light Seed Black Holes in the Early Universe}

\author{\textbf{$^*$Daxal ~H. Mehta$^{1,2}$, John ~A. Regan$^{1,2}$, Lewis Prole$^{1,2}$}}

\begin{document}

\maketitle

Light seed black holes (LSBHs) form as the end point of the first generation of Population III (PopIII) stars in the early universe, forming as early as $z \sim 30 - 40$ \cite{madau2001massive, volonteri2003assembly, madau2004early, Volonteri_2012} in low-mass dark matter halos. PopIII stars form from metal-free gas, are expected to be more massive \cite{Latif_2022} than subsequent generations of metal-poor stars, and exhibit a mass range from sub-solar to hundreds of solar masses \cite{abel2002formation, bromm2002formation, o2007population, turk2009formation, hirano2014one, prole2023dark}.  Depending on their zero age main sequence (ZAMS) mass, PopIII stars transition into black holes (BHs) via either the supernova (SNe) or direct collapse channel \cite{woosley1995evolution, nomoto2006nucleosynthesis, heger2002nucleosynthetic}. The large number density of LSBHs in the early universe makes them ideal candidates to be the progenitors of the observed SMBHs, as only a small fraction of the overall population needs to grow to explain the observed abundance of SMBHs. LSBHs can grow into SMBHs if they can either sustain near-Eddington accretion for Gyr time periods, or more likely experience short bursts of super-Eddington accretion \cite{Lupi_2016, smith2018growth, Regan_2019,sassano2023super}. In this vein, many theoretical studies argue that the formation of a thick disk around a BH may facilitate super-Eddington accretion even in the presence of feedback \cite{inayoshi2016hyper, jiang2019super, park2020biconical, kitaki2021origins, botella2022structure}. This pathway is further supported by recent observations of SMBHs that accrete at super-Eddington limits \cite{lambrides2024case, suh2025super}.\\
\indent However, even if LSBHs can accrete at super-Eddington rates, it is still unclear whether LSBHs encounter regions of sufficiently high density to allow super-Eddington accretion to develop efficiently and frequently. This is particularly true at mass scale of the LSBHs since the lighter a BH is, the more difficult it is to attract mass and accrete. Previously, it has been shown that LSBHs can grow in idealised, isolated galaxy simulations, using extremely high resolution hydrodynamic simulations\cite{shi2023hyper, gordon2024hungry, mehta2024growth}. Depending on the properties of the surrounding gas, it was shown that LSBHs can grow on timescales of 500 Myr \cite{Lupi_2016} or even achieve rapid runaway accretion episodes \cite{Alexander_2014, shi2023hyper, mehta2024growth}. Several studies have shown that super-Eddington accretion onto light black hole seeds is possible across a wide range of parameters, including feedback efficiencies, seed masses, and accretion models, and that sustained growth can occur even when thermal feedback is included \cite{gordon2024hungry, zana2025super}. However, more realistic, large-scale cosmological simulations have consistently failed to demonstrate LSBH growth \cite{Alvarez_2009, smith2018growth}. 

\begin{table*}
    \centering
    \renewcommand{\arraystretch}{1.2} 
    \begin{tabular}{|>{\centering\arraybackslash}p{2.0cm}  
                    |>{\centering\arraybackslash}p{2.1cm}  
                    |>{\centering\arraybackslash}p{2cm}    
                    |>{\centering\arraybackslash}p{3.2cm}  
                    |>{\centering\arraybackslash}p{2.5cm}  
                    |>{\centering\arraybackslash}p{2.2cm}|}
        \hline
        \textbf{Simulation} & \textbf{Boxsize [Mpc h$^{-1}$]} & \textbf{M$_{DM}$ [M$_{\odot}$ h$^{-1}$]}& \textbf{Gravitational Softening [pc]} & \textbf{Min cell length [pc]} & \textbf{Final Redshift} \\
        \hline
        L13 & 1 & 10177 & 47 & 1 & 18 \\
        \hline
        L14 & 1 & 1270 & 15 & 0.3 & 18.8 \\
        \hline
        L15 & 0.5 & 159 & 6 & 0.1 & 21 \\
        \hline
        L15\_BHFB & 0.5 & 159 & 6 & 0.1 & 21 \\
        \hline
    \end{tabular}
    \caption{\textbf{Details of the simulations:} \textit{C1: Simulation name, C2: Boxsize of the simulation (comoving Mpc h$^{-1}$). C3: Mass resolution of dark matter particle (\msolar h$^{-1}$). C4: Softening length of dark matter particle (physical pc). C5: Softening length and minimal spatial resolution of gas cell (physical pc). C5: Final redshift the simulation reached.\\}}
    \label{tab:simulations}
\end{table*}

\indent In this study, we run a suite of high-resolution cosmological simulations to investigate the small spatial scale dynamics (sub-pc) around LSBHs and study their growth in a fully cosmological setting, including both SNe feedback and BH feedback. Our goal is to investigate if LSBHs (similar to their behaviour in idealised isolated systems) can grow in a self-consistent cosmological environment. We use the moving-mesh code Arepo \cite{Springel_2010, Pakmor_2016} to perform all simulations in this study. We use MUSIC \cite{hahn2011multi} to generate three sets of initial conditions at progressively higher levels of mass and spatial resolution (named L13, L14 and L15, see Table \ref{tab:simulations} for more information). For each of these resolutions, we use the same set of random number seeds to give near identical distribution of structures in each simulation box. This simulation suite is part of the larger \textsc{SEEDZ} simulation suite whose goal is to investigate the origin of SMBH seeds \cite{mehta2024growth, Prole_2025, Prole_2025b} . Both simulation suites include models of metal-free star formation (PopIII), metal-poor star formation (PopII), metal enrichment, SNe feedback, BH accretion, BH feedback and BH mergers. As in our previous work \cite{Prole_2025}, we also include a redshift-dependent Lyman-Werner (LW) background flux to mimic the feedback from nearby star-forming regions, which (theoretically) lie outside of our simulation volume \cite{qin2020tale}. \\
\indent For our lowest-resolution simulation (L13), the minimum spatial resolution is 1\,pc (physical), and the minimum particle mass resolution is $10,\!177\, \rm{M_\odot\,h^{-1}}$. In the higher-resolution simulations, L14 and L15, the minimum spatial resolution improves to 0.3\,pc and 0.1\,pc, respectively, with corresponding particle mass resolutions of $1,\!270\,\rm{M_\odot\,h^{-1}}$ and $159\,\rm{M_\odot\,h^{-1}}$. Finally, realisation L15\_BHFB has the same spatial and mass resolution as the L15, but also includes thermal feedback due to accretion onto the BHs.  In the L15 simulation, the dark matter particle mass is extremely well resolved (159 \msolar $\rm{h^{-1}}$) - similar to that of a massive PopIII star, which means that for our highest resolution simulations, the gravitational dynamics are highly resolved. Perhaps even more importantly, at this resolution, we are able to resolve the Bondi-Hoyle-Littleton (BHL) radius of the smallest black holes, enabling us to robustly capture the accretion around these BHs.\\
\indent The simulations were initialized at $z = 127$ and were evolved to $z \sim 21$ ($z \sim 18$ for the lower resolution runs). PopIII stars began forming in each realisation at $z > 30$. Due to the relatively short lifetime of massive PopIII stars ($\sim 2$ Myr), LSBHs begin to form almost immediately after PopIII star formation commences. In each realisation, the stellar masses of the PopIII stars are sampled from a top-heavy IMF with a characteristic mass of 20 \msolar (see extended data Figure 1). In total, we form 1770, 4318, 1672, and 1015 PopIII stars in L13, L14, L15, and L15\_BHFB respectively (note that L15 and L15\_BHFB have smaller volumes for reasons of computational efficiency). The decrease in the number of PopIII stars in L15\_BHFB  relative to L15 is because thermal feedback from accretion regulates star formation. In total, these runs produce 6827, 16286, 1565, and 1527 halos, out of which 326, 403, 64, and 64 go on to form galaxies and host BHs, respectively (see extended data Figure 2).\\
\indent Upon creation, each PopIII star is assigned a stellar lifetime according to its ZAMS mass, after which they transition into stellar mass BHs, undergoing a direct collapse or SNe (again depending on their assigned ZAMS mass \cite{ woosley1995evolution, heger2002nucleosynthetic, nomoto2006nucleosynthesis} - see for example extended data Figure 3). Once formed, BHs can start accreting the gas surrounding them. We model this feature of BH growth using the classical Bondi-Hoyle-Lyttleton (BHL) method. Further details of the sub-grid models can be found in our Methods section.\\
\begin{figure}
    \centering
    \includegraphics[width=\linewidth]{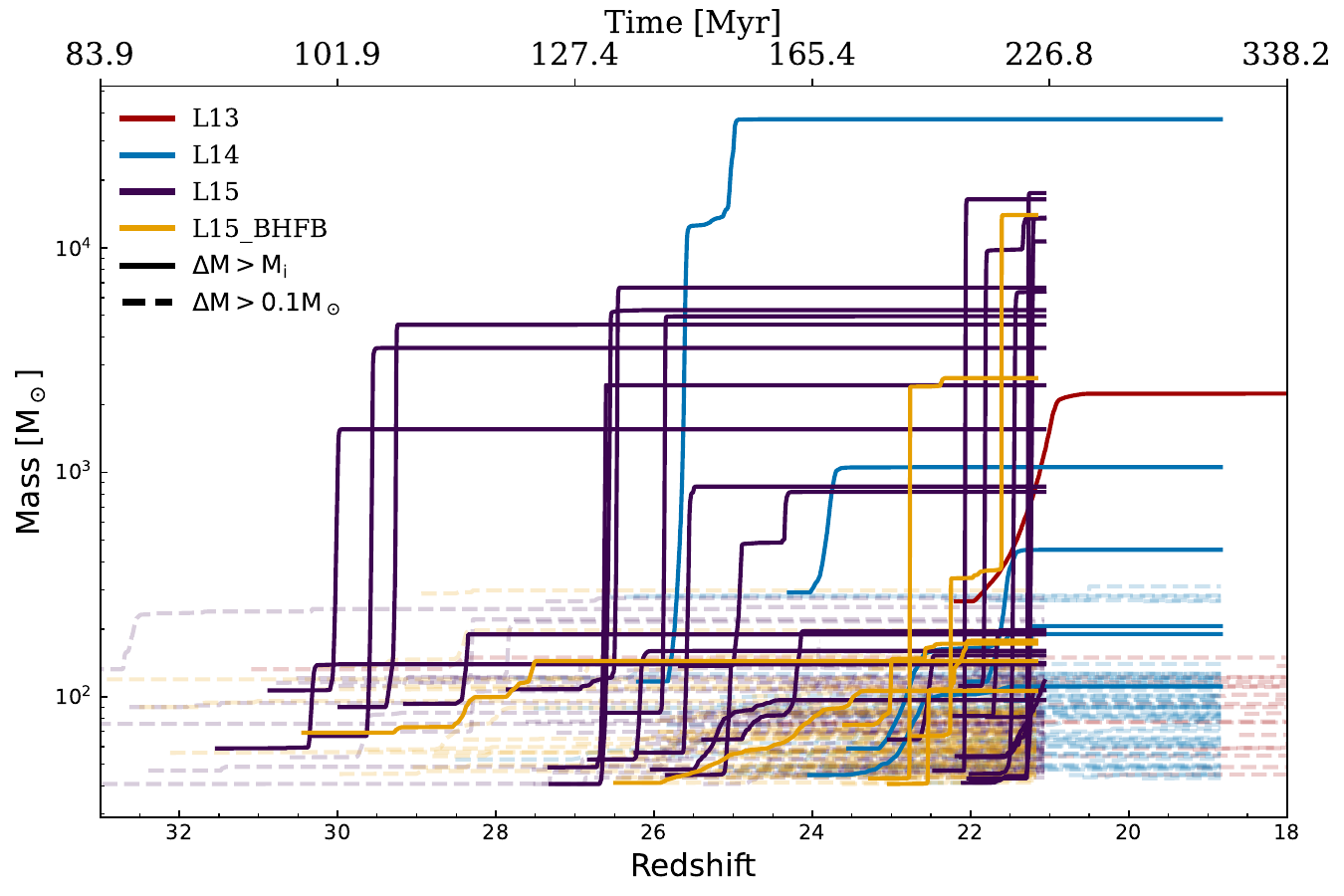}
    \caption{\textbf{Mass growth history of BHs:} We show all BHs that accreted more than 0.1 \msolar (dashed lines) and all BHs that doubled their initial mass (solid lines). The simulations L13, L14, L15 and L15\_BHFB are coloured red, blue, violet and yellow respectively. Firstly, we see that the number of growing BHs increases with resolution.  Secondly, we see that the BHs accrete very rapidly and their rapid-accretion episodes are often short-lived, lastly at most a few million years. In this time, many BHs are able to realise super-Eddington accretion levels with many BHs growing to more than $10^4$ \msolar in only a few Myr.}
    \label{fig:StellarMassGrowth}
\end{figure}
\begin{figure*}
    \centering
    \includegraphics[width=\linewidth, height=13cm]{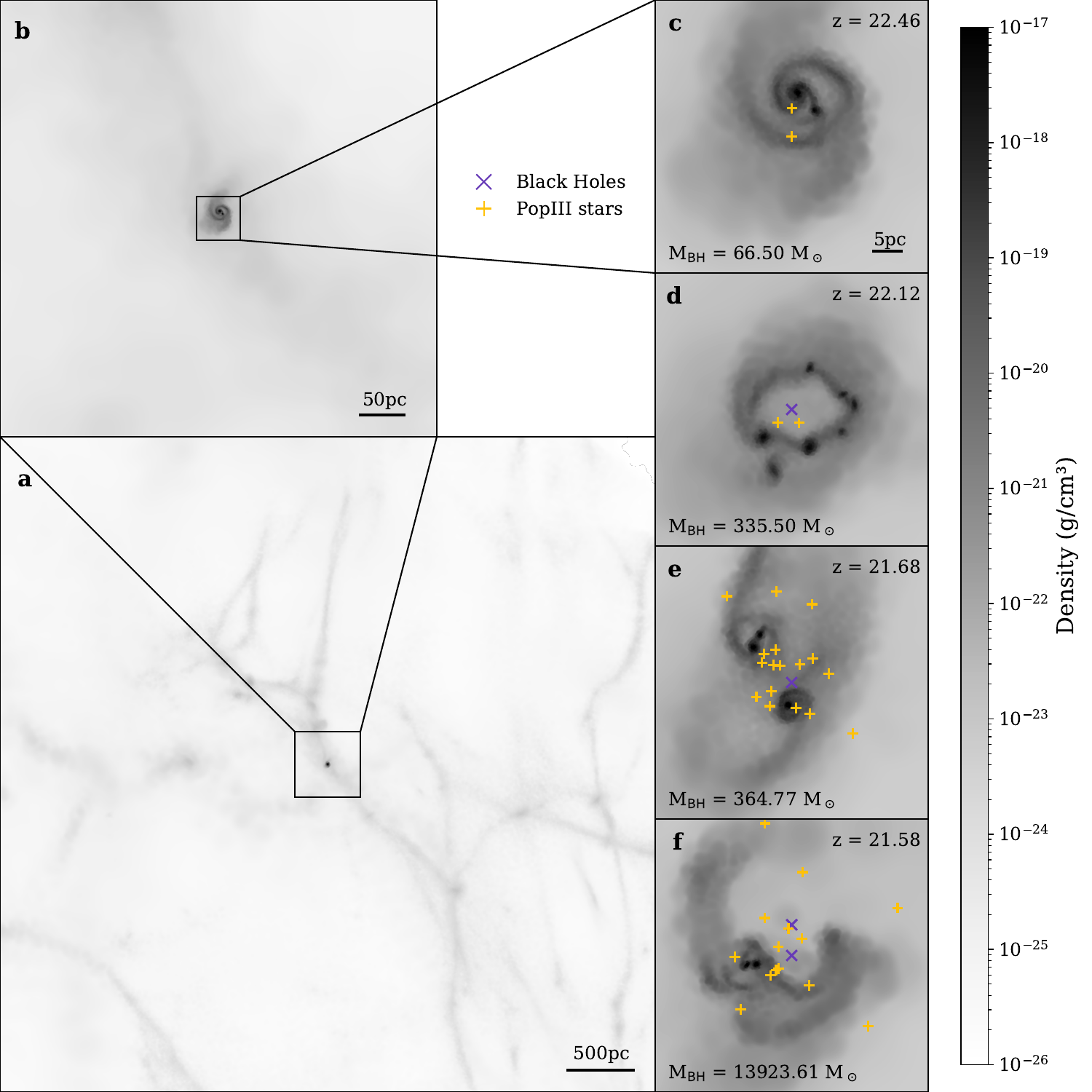}
    \caption{\textbf{Gas density projection for PopIII star formation, BH formation, supernova feedback and thermal feedback for the most massive BH in the L15\_BHFB simulation:} a, b, c: Within cosmological filaments, small mini-halos collapse into galaxies and begin metal-free PopIII star formation. The galaxy forms a disk like structure with clear spiral arms. d: The first PopIII transitions to a BH through direct collapse and accretes mass, reaching 335 \msolar. The injected thermal energy causes the galaxy to distort in the centre. e: The galaxy reassembles and collapses back onto the BH, triggering a period of intense star formation. f: SNe feedback from another PopIII star injects significant energy into the galaxy, once again distorting it. Eventually, other PopIII stars undergo SNe combined with the thermal BH feedback completely expel the gas from the galactic centre.}
    \label{fig:gas_visualizations}
\end{figure*}
\indent In Figure \ref{fig:StellarMassGrowth}, we show the mass growth of all BHs who accrete more than 0.1 \msolar (dashed lines) and all BHs that doubled their initial masses (solid lines). Growth in all cases is primarily via gas accretion. Mass growth by BH mergers is negligible in all cases. We only see eight BH mergers across all simulations representing a negligible contribution to the overall BH growth.\\
The highest-resolution runs (L15 and L15\_BHFB) show the largest number of growing BHs, despite having volumes eight times smaller than the other runs. This underscores the importance of resolving the BHL radius ($\sim$ 0.5 pc for a 100 \msolar BH) to robustly capture the BH growth trajectory. These two simulations achieve a spatial resolution of 0.1 pc, where we can better model the complex gas flows leading to accretion. Simultaneously, the mass resolution accurately captures the dynamics of BHs in the background sea of stars and dark matter particles. As a result, within our highest-resolution runs, we see the growth of numerous BHs to in excess of $10^4$ \msolarc. We here demonstrate LSBH growth in a full cosmological setting, including the impact of both SNe and BH (thermal) feedback.\\
\indent Quantitatively, we identify 1(0.05 \%), 6(0.14 \%), 25 (1.5 \%), 6 (0.6 \%) BHs in the L13, L14, L15 and L15\_BHFB simulations that experience growth of more than 100\% of their initial seed mass, with the most massive BH reaching 2230, 37205, 17554, and 13923 \msolar, respectively. In the case of L15\_BHFB, the feedback from BH accretion sharply decreases the number of rapidly accreting BH compared to the L15 simulation. The (thermal) feedback evacuates the dense gas around the BH, stunting further growth in the majority of cases. Nevertheless, several BHs still achieve noticeable growth, either before the feedback fully evacuated the nearby dense gas or in some cases the gas re-collapsed onto the BHs, driving further growth. Despite our strong feedback model, (see our Methods section), we still see the formation of a BH at almost $1.4 \times 10^4$ \msolar. showing that LSBHs can experience significant growth even with the pressures of negative feedback. \\ 
\indent In Figure \ref{fig:gas_visualizations}, we examine the growth of the most active BH from the L15\_BHFB realisation which is our highest resolution simulation and includes BH thermal feedback. This BH reaches a final mass of approximately 14,000 \msolar starting from an initial mass of just 66 \msolarc. In Figure \ref{fig:gas_visualizations}-a, b \& c we show the initial halo in which the progenitor PopIII star forms. Panel $a$ shows the large scale structure, panel $b$ shows a zoom-in onto the host halo and panel $c$ shows a further zoom centred on the PopIII progenitor star and spiral-like structure of the embryonic galaxy. Panel $d$ captures the transition of the PopIII star into a BH (via direct collapse) and the beginning of accretion of surrounding gas onto the BH up to $\rm{M_{BH}} \sim 335$ \msolarc. This is followed by the release of a significant amount of thermal energy from the growing BH that disturbs the host spiral galaxy and creates a central cavity halting accretion. In panel $e$ we observe further gas infall and the subsequent formation of a multitude of additional PopIII stars. The infalling gas also feeds the BH, restarting accretion and allows rapid growth of the BH up to almost 14,000 \msolar - despite the impact of thermal feedback from the BH. However, the thermal feedback coupled with SNe feedback from the surrounding PopIII stars eventually halt further accretion with a few Myr (panel $f$). Nonetheless, it is remarkable that the LSBH has reached a final mass (by z $\sim 21.5$, also see extended data Figure 4 for temperature projections) of almost 14,000 \msolar despite the obvious negative feedback effects of both accretion feedback and SNe feedback, agreeing with previous idealized works \cite{shi2024feedback, gordon2025conditions, zana2025super}.\\
\begin{figure}
    \centering
    \includegraphics[width=\linewidth]{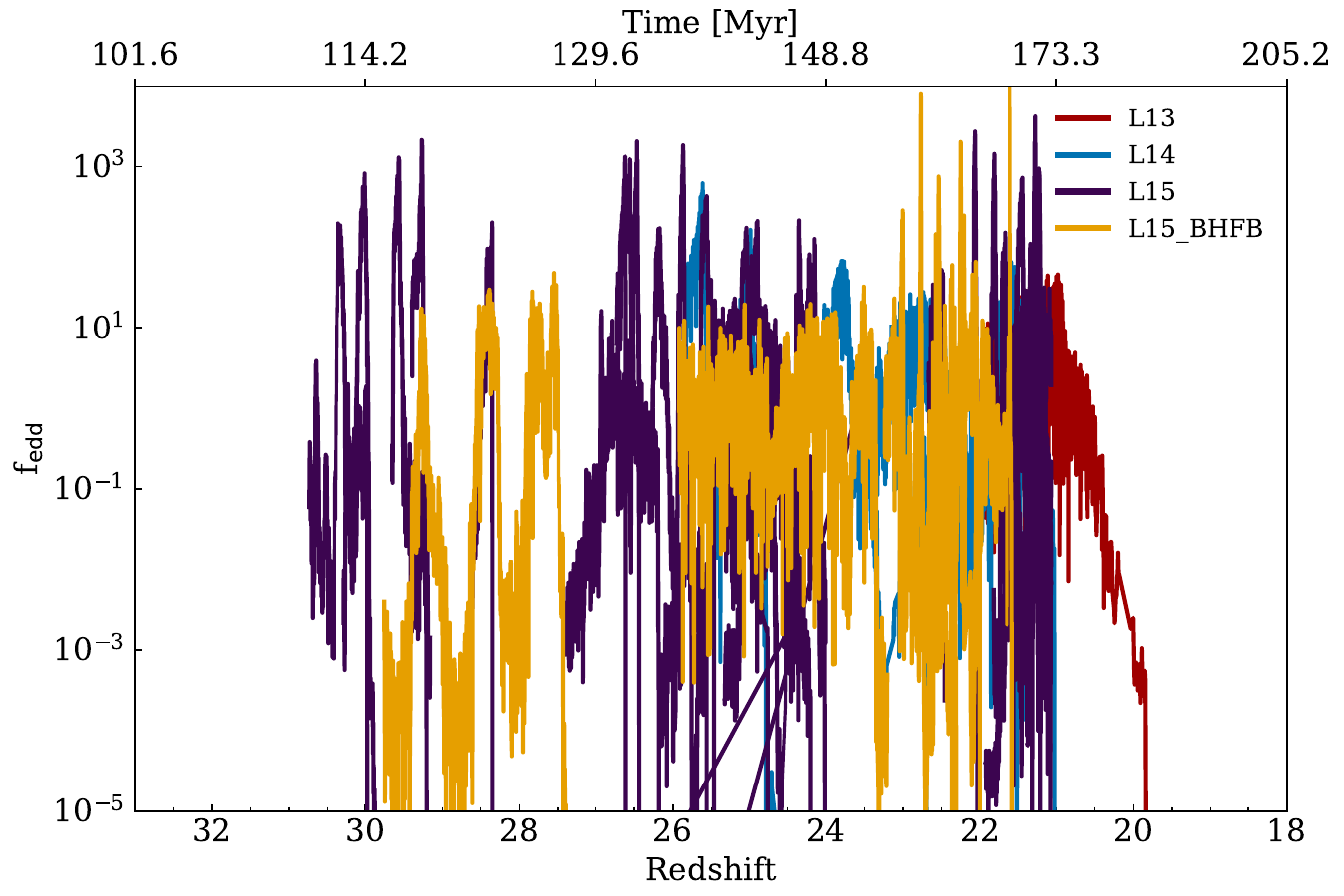}
    \caption{\textbf{Black Hole growth in terms of the Eddington fraction:} We show the Eddington factors of all BHs which are growing sufficiently to double their mass. The simulations L13, L14, L15 and L15\_BHFB are coloured red, blue, violet and yellow respectively. We see that regardless of resolution, BHs are able to achieve accretion rates in excess of the canonical Eddington limit. For several time periods in our highest resolution simulations maximum accretion rates can exceed Eddington factors of $10^3$ albeit only for very short durations. Nonetheless, this clearly demonstrates that super-Eddington accretion is possible and indeed likely a necessary condition to drive rapid growth at high redshift.}
    \label{fig:EddingtonRate}
\end{figure}
\begin{figure*}
    \centering
    \includegraphics[width=\linewidth, height=13cm]{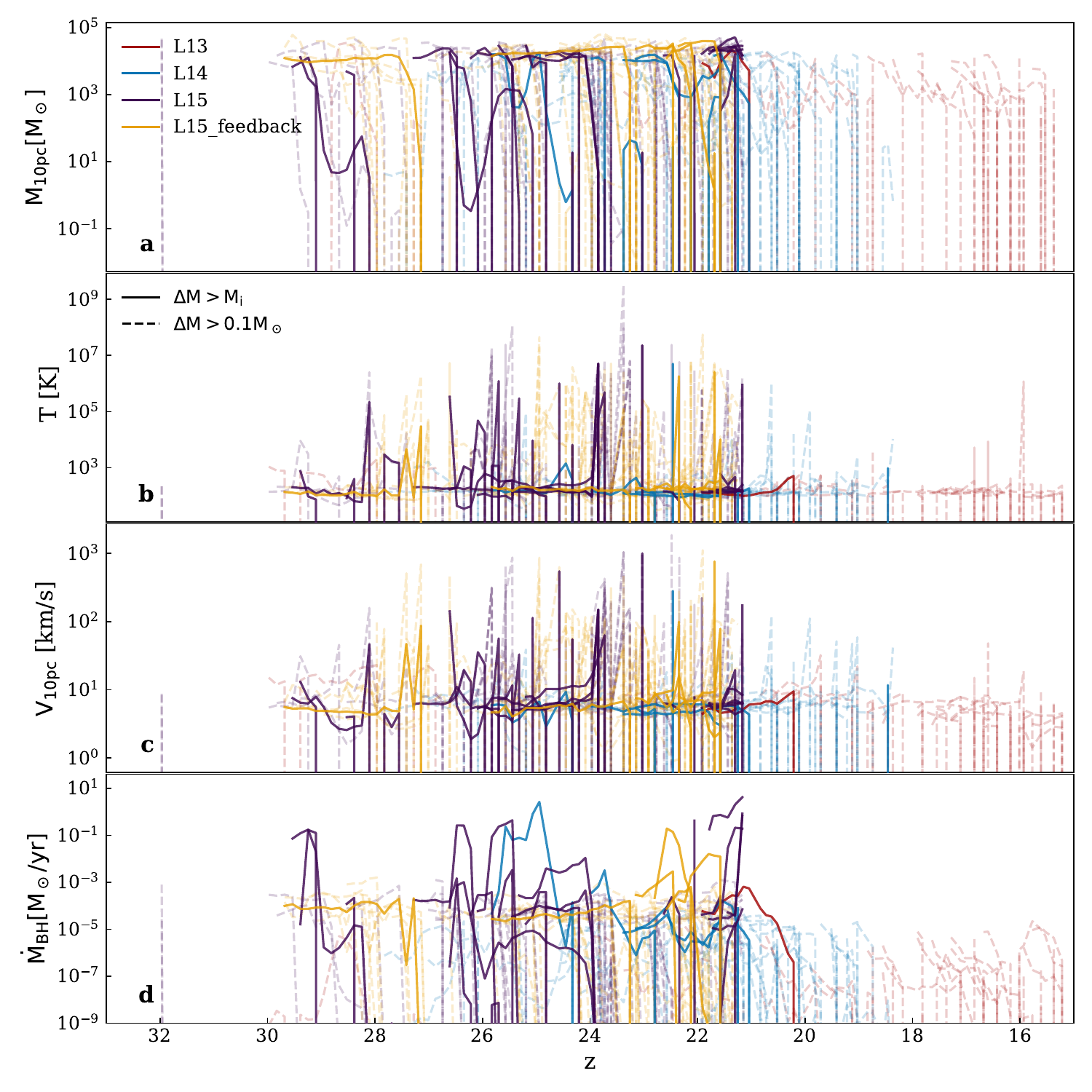}
    \caption{\textbf{Gas properties within 10 pc of growing BHs.} We show the density-weighted gas properties within a 10 pc sphere surrounding all BHs that grew more than 0.1 \msolar (dashed) and for BHs that double their initial seed mass (solid lines). The simulations L13, L14, L15 and L15\_BHFB are coloured red, blue, violet and yellow respectively. In (a), we show the total mass of the gas surrounding the BHs as the BHs evolve with redshift. We notice that for almost all BHs, there is a drastic decrease in mass as rapid accretion begins. This is because SNe feedback expel gas from the mini-halos, raising the gas temperature beyond $10^5$ K (b) and driving relative radial velocities to 1000 km/s (c), higher than the escape velocities of the mini-halos. All of these factors cause the accretion rate onto the BHs to sharply decrease (d), completely stopping accretion onto the BHs.}
    \label{fig:SurroundingGasProperties}
\end{figure*}
\indent The BHs that do grow in our simulations, accrete rapidly within timescales of $10^5-10^6$ yr (see Figure \ref{fig:StellarMassGrowth} and extended data Figure 5). The growth often exceeds the canonical Eddington rate leading to rapid bursts of super-Eddington growth and feedback. In Figure \ref{fig:EddingtonRate}, we show the Eddington factors of the BHs that doubled their seed mass. Mean accretion rates are clearly close to Eddington but with bursts that can exceed the Eddington rate by up to a factor of 1000. Even short duration bursts of super-Eddington accretion can lead to extremely rapid growth and the formation of a BH with a mass in excess of $10^4$ \msolar and hence the rapid formation of a so-called intermediate mass black hole. \\
\indent Ultimately, BH growth is shut down in all of our cases by external environmental impacts - either SNe feedback, BH feedback or gas starvation. As show in Figure \ref{fig:SurroundingGasProperties} (panel \textit{d}), we see high accretion rates onto the growing BHs ($\gtrsim 10 ^{-3}$ \msolar$/\rm{yr}$), when the BHs are surrounded by cold $(\rm{T} \lesssim 10^3$ K, panel \textit{b}) and slow moving ($\rm{V_{rel, 10pc}} \lesssim 10 \ \rm{km s^{-1}}$, panel \textit{c}) gas in sufficient quantities (panel \textit{a}). However, SNe feedback from nearby stars and potentially from the BH itself inject significant energy and momentum into the interstellar medium surrounding the BHs, heating the gas to temperatures exceeding $10^5$ K and driving gas (relative) velocities over 1000 km/s. This causes the accretion rate to rapidly decline. The gas is evacuated from the BH's surroundings and accretion can only resume when the gas re-collapses into the vicinity around the BHs. However, this is not guaranteed and in the majority of cases the growing BHs had a singular growth phase and subsequent phases were not observed in our simulation timescales. Additionally, for some BHs, we see the accretion rates decline solely because the BH accreted all of its surrounding gas. These BHs become starved until such time as they can find another cold and dense gas clump.

\indent LSBH growth is not only possible in pristine metal-free galaxies but also in highly metal enriched environments as gas re-collapses and growth can be triggered again and again. A fraction of BHs grow in pristine metal-free conditions (extended data Figure 6) but a large fraction also grow in galaxies enriched by metals from SNe, with metal enrichment levels typically in excess of $10^{-4} Z_{\odot}$. In fact, the L14 and L15 simulations have BHs growing larger than $10^4$ \msolar in metal-enriched galaxies. \\
\indent We also investigate the correlation between the growth of BHs  with the properties of their host halos (see extended data Figure 7). Examining the final BH masses as a function of host halo mass at the time of PopIII star formation reveals no discernible trend. Instead, we see that the host halo generally exceeds $10^6$ \msolar but is not necessarily heavier than the atomic cooling threshold (approximately $10^7$ \msolar at z = 24) i.e. only a halo capable of supporting star formation is required which is of course a minimum requirement. We similarly see no direct correlation between BH growth and the surface density of its surrounding gas (panel $b$ of extended data Figure 7). In our simulations, once the surface density exceeds a minimum value of $\sigma \gtrsim 20$ \msolar \ pc$^{-2}$, rapid growth is possible. \\ 
\indent An interesting, albeit not totally surprising, pattern emerges when comparing the final masses of the BHs to the initial masses of their PopIII progenitors (see extended data Figure 3). Almost all growing LSBHs were formed through the direct collapse channel. We see no growth for BHs that formed through the SNe channel. This contrast suggests that SNe feedback effectively clears the local environment of gas, suppressing further accretion, while the direct collapse pathways—lacking such feedback—enable rapid early growth in high-density regions. This result correlates with previous works in the field which showed that LSBH initially struggle to grow\cite{Alvarez_2009, smith2018growth}. We note here that our simulations do not include the potentially negative feedback effects from the PopIII progenitor star phase. However, the impact of radiative feedback from PopIII stars is not expected to be significant within the dense confines of high-z galaxies \cite{Regan_2020b, jaura2022trapping}. Instead, high resolution numerical studies have shown that the ionizing radius is instead trapped within a small HII region around the star preventing large scale breakout \cite{Regan_2020b, jaura2022trapping}. \\
\indent So then what does determine the probability of a BH to grow? From our simulations we find that the growth is stochastic and occurs in bursts. There appears no clear correlation between BH growth and the host halo macro environment. Instead, small scale fluctuations are likely to be more important. Firstly, BHs born via the direct collapse channel have an obvious advantage in that an initial SNe explosion does not evacuate the surrounding material. Secondly, more than one episode of growth may be required due to the negative feedback effects of BH accretion. However, once a light seed black hole undergoes a phase of rapid early growth, it gains several advantages for subsequent evolution. Its radius of influence increases significantly with mass, enhancing its ability to gravitationally bind surrounding gas and stars. In addition, the larger dynamical mass strengthens the effect of dynamical friction by a small fraction, improving the likelihood of the black hole migrating toward the galactic centre.\cite{Tremmel_2015, Pfister_2019}.\\
\indent Interesting, the final masses of our growing light seeds $10^3-10^4$ \msolar are the seed BH masses in larger cosmological simulations \cite{taylor2014seeding, habouzit2017blossoms, bhowmick2024introducing}. Many of these studies found that their seeds grow into SMBHs at lower redshifts and are likely to be progenitors of the SMBHs observed at lower redshifts. But for these seeds to sustain growth, their host galaxies stellar mass must typically exceed masses of $~10^{9}-10^{10}$ \msolar at which point feedback is no longer efficient at expelling gas. Our simulations therefore provide a bridge between the light seeds expected from the first stars and the black holes seeded in large-volume cosmological simulations, linking the earliest formation pathways to the long-term growth into the SMBHs observed at low redshift. \\
\indent Our findings clearly demonstrate that LSBHs can undergo sufficient growth in a cosmological setting, provided the simulations have sufficiently high resolution to resolve the BHL-radius of the LSBHs. This, of course, is a non-trivial task in the mainstream of cosmological simulations. PopIII remnant black holes are typically not modelled in a cosmological context due to the high resolution required to resolve their birth environments and secondly the required resolution to correctly model the growth of their remnants is typically not achieved (i.e. the BHL radius is not resolved). Our simulations achieve both but at the cost of being unable to evolve the system beyond a few hundred Myr post Big Bang. \\
\indent Although our high-resolution simulations are limited in runtime and cannot trace the long-term evolution of these seeds, the early growth that these select few BH undergo positions them as compelling candidates to be the dominant channel for the origin of SMBHs. The number density of LSBHs which reach masses in excess of $1000$ \msolar is large enough, $> 10 \ \rm{cMpc}^{-3}$ (approximately 10 times larger for the realisations without BH feedback) at $z = 21$ in our simulations (see extended data Figure 8) that we only need a tiny fraction of these to encounter further suitable conditions to match the current high-z SMBH population of candidate AGN( $\lesssim  10^{-2} \ \rm{cMpc}^{-3}$) \cite{perez2024nature, kokorev2024census}. \\

\subsection{References}
\bibliographystyle{naturemag}

\noindent \textbf{Data Availability:}   The simulation outputs generated and analysed in this study amount to approximately 6 TB and cannot be hosted in a public repository due to their size. These data are available from the corresponding author upon reasonable request, and we will provide the full set of snapshots and derived data products necessary to reproduce the analyses presented in the paper. Summary products required for figure generation (BH growth histories, gas properties, halo catalogues, and extracted time series) have been deposited in a public repository at figshare \cite{datafigshare}

\noindent \textbf{Code Availability:}   The simulations were carried out with a proprietary version of the publicly available AREPO code. The publicly released version of AREPO can be obtained from the Max Planck Institute for Astrophysics (https://arepo-code.org). The modified, proprietary version used for our production runs cannot be redistributed. All analysis scripts developed for this study—including routines for processing the snapshots, computing derived quantities, and generating the figures—are available at a figshare repository \cite{codezenodo}

\begin{addendum}
 \item 
J.A.R acknowledges support from the Royal Society and Research Ireland under grant number URF/R1/191132. D.H.M, J.A.R and LP acknowledge support from the Research Ireland Laureate programme under grant number IRCLA/2022/1165. The simulations were performed on the Czech Republic EuroHPC machine Karolina hosted by IT4Innovations through a EuroHPC Regular Access call (EHPC-REG-2023R03-103) and on the Luxembourg machine Meluxina. The authors wish to acknowledge the Irish Centre for High-End Computing (ICHEC) for the provision of computational facilities and support. The authors also acknowledge the constructive comments from the referees.
 
 \item[Author contributions] D.H.M, J.A.R, and L.P conceived the idea of the project. D.H.M performed the simulation, analysis and drafted the paper. J.A.R performed the coarse 40 Mpc simulations. All authors contributed to the interpretation of the results and to the text of the final manuscript.
 \item[Competing Interests] The authors declare that they have no competing financial or non-financial interests.
 
 \item[Materials \& Correspondence] Correspondence and requests for materials should be addressed to D.H.M. ~(email: daxal.mehta.2024@mumail.ie).
\end{addendum}

\newpage
\begin{methods} \label{Methods}
\subsection{Cosmological Setup\\}
\label{sec:setup}
The simulations in this study were carried out using a moving mesh refinement code Arepo \cite{Springel_2010, Pakmor_2016}, which solves the Euler equations on an unstructured Voronoi tessellation moving with the gas flow. The fluid dynamics are solved using a finite-volume, second-order reconstruction scheme coupled with an exact Reimann solver. We employ the hierarchical time integration scheme \cite{Springel_2021}, which splits the Hamiltonian into "fast" and "slow" components, i.e. particles on short versus long time-steps, respectively. With this approach, interactions between particles/cells on the smallest time steps are solved using direct summation, while a standard octree approach is used for all other interactions. This is primarily done for efficiency, to avoid unnecessary tree constructions for small number of particles, but has the added benefit of more accurately calculating the gravitational forces between BHs and their nearest neighbours, which are on the short time-steps, via direct summation \cite{bourne2024dynamics}. Refinement of gas cells is based on the Jeans refinement criteria and is done to ensure that the Jeans length of the gas is always refined by at least four cells. The refinement is halted once the minimum cell size is reached (see Table \ref{tab:simulations}). De-refinement is also employed, so that gas expanded by feedback events adaptively increases its cell size. To avoid sharp discontinuities, neighbouring gas cells are restricted to sizes within a factor of three of each other. A maximum cell size of 1000 ckpc/h is permitted, though this limit is rarely, if ever, reached in practice because of the neighbouring cell condition.

\subsection{Initial Conditions \\}
The simulations presented here were carried out in 2 stages. The parent simulation was initialized at $z=127$ using initial condition generated from MUSIC \cite{hahn2011multi} with a co-moving box of side 40 cMpc h$^{-1}$. We use the standard $\Lambda$CDM cosmology parameters $h = 0.6774$, $\Omega_0 = 0.3089$, $\Omega_b = 0.04864$, $\Omega_\Lambda = 0.6911$, $n=0.96$, and $\sigma_8=0.8159$ \cite{Planck2020}. The parent simulation is dark matter only and is evolved until $z=10$, where we use Friend-of-friends (FOF) algorithm to identify the largest halo in the box, using it as the centre for our zoom-in simulations. Using the same set of random number seeds, we recreate the ICs centred on the halo with box size 1 cMpc h$^{-1}$. For the L15 and L15\_BHFB simulations, we further decrease the box size to just 0.5 cMpc h$^{-1}$. In the zoom-in simulations, we split the dark matter particle to form gas particles according to the cosmic baryonic fraction.

\subsection{Star Formation\\}
\label{sec:star_formation}
\indent The star formation technique employed in this work builds and improves on the scheme originally introduced into Arepo \cite{wollenberg2020formation, tress2020simulations}. Star particles are formed in high density gas regions when a gas cell can no longer refine upon reaching the minimum cell size of the mesh. They are point-like particles artificially introduced to model the behaviour of physical processes that occur below the resolution of our mesh. The approach taken here is broadly similar with the criteria described in Krumholz et al. (2004)\cite{krumholz2004embedding} and previously employed in the ENZO code \cite{Enzo2014, regan2018rise, Enzo2019}. In this study, we adapt the formation criteria so that a cell is converted into a PopIII particle after satisfying the first three conditions, 

\begin{enumerate}

    \item The density of the gas cell should be higher than the local Jeans density defined by the criterion \cite{truelove1997jeans},
        \begin{equation}
            \rho > \rho_J = J^2 \frac{\pi c_s^2}{G\Delta x^2},
            \label{eq:Jeans_density}
        \end{equation}
        where $\Delta x$ is the cell size and $J=0.25$ is a constant \cite{truelove1997jeans}.
    \item The cell is sufficiently far away from a pre-existing star. The gas cell should be outside a sphere of 5 times the accretion radii of other star particles. This exclusion zone ensures that we do not get spurious star formation or spurious BH mergers.
    \item The gas within this region is gravitationally bound and collapsing.
    \item In addition to satisfying the above conditions, if the gas cell has metallicity higher than $10^{-4}$ Z$_\odot$, we form a PopII star cluster particle instead of a PopIII particle.
\end{enumerate}

\subsection{Assigning Star Particle Masses\\}
For the PopIII star particles, the stellar mass is randomly sampled from a top-heavy initial mass function (IMF).
\begin{equation}
    \rm{f(log M) dM = M^{-1.3} exp \Big[ \big( \frac{M_{char}}{M} \big)^{1.6} \Big] dM}
\end{equation}

The IMF is constructed with a lower and upper bounds at 1 \msolar and 300 \msolar, respectively, and the slope is -1.3. The characteristic mass $\rm{M_{char}}$ is set to be 20 \msolar. A stellar lifetime is also assigned to the PopIII particle, depending on its stellar mass \cite{maeder2008physics}. If the sampled stellar mass is higher than the host gas cell mass, we do not form a PopIII particle.

For PopII star cluster particles, we assume a star formation efficiency of $\epsilon = 0.1$, forming a cluster particle of mass $\epsilon M_{host}$, where $M_{host}$ is the mass of the host gas cell.  The total cluster mass is then populated with individual PopII stars sampled iteratively from a Kroupa IMF in the mass range 8-104 \msolar, until the cluster mass is fully assigned. We do not need to sample the stars below 8 \msolar as these stars will have lifetimes longer than our typical simulation timescales and hence we not need to identify them. However, these stars exist as part of the PopII cluster and hence only a small portion of the PopII cluster is converted back into gas and metals as part of the SNe process. As with PopIII stars, each PopII star, above 8 \msolarc, is given a stellar lifetime, after which it undergoes a supernova event. 

\subsection{Supernova Feedback\\}
\label{sec:supernova}
PopIII stars and individual PopII stars undergo a SNe after their stellar lifetimes have passed. The various types of SNe and their energy output are as follows:
   \begin{enumerate}
      \item Star particles between 11-20 \msolar undergo a type II SNe, with energy $E_{SN}$=$E_{51}$=$10^{51} $ ergs. The helium core mass post-explosion is calculated as \cite{nomoto2006nucleosynthesis}:
      \begin{equation}
    m_{\rm He} = 0.1077 +  0.3383 (m_{\star}-11),
    \label{eq:He_core_small}
    \end{equation} 
    where $m_{\star}$ is the stellar mass. 
    \item Star particles in the mass range of 20-40 \msolar undergo a hypernova with helium core mass interpolated from a fitting formula \cite{nomoto2006nucleosynthesis}.
    \item Star particles in the mass range of $40-140$ \msolar experience a direct collapse into a BH and exhibit no SNe explosion. 
    \item Star particles (only PopIII in this case) between 140 - 260 \msolar undergo a PISN, with the explosion energy and helium core mass as:
      \begin{equation}
      m_{\rm He} = \frac{13}{24} (m_{\star} - 20)
      \label{eq:He_core_large}
      \end{equation}
      \begin{equation}
      E_{\rm SN} = 5 + 1.304 (m_{\rm He} - 64).
      \label{eq:SN_E51}
      \end{equation}
    These star particles do not convert into a BH and are instead deleted from the simulation.
    \item Lastly, star particles (only PopIII) heavier than 260 \msolar convert into a BH through the direct collapse channel. No SNe occurs.
    \end{enumerate}
\indent Unlike extremely high resolution works that ensure the Sedov-Taylor scale is resolved to accurately inject SNe energy into the simulation mesh \cite{gatto2015modelling, tress2020simulations, magg2022metal}, we instead opt to model the explosion as an injection of momentum. The momentum is injected into the nearest 1000 cells to the star particle. We calculate the momentum of the SNe depending on the SNe energy, the density and metallicity of the injection region\cite{smith2019cosmological}. 
\begin{equation}
    p_{\rm SN} = 3.0 \times 10^5 E_{\rm 51}^{16/17} n_{\rm SN}^{-2/17} Z_{\rm SN}^{-0.14} \rm{M_\odot} \rm{km/s},
    \label{eq:mom_inj}
\end{equation}
where $E_{\rm 51} = E_{\rm SN}/10^{51}$, $n_{\rm SN}$ is the number density of the injection region, and $Z_{\rm SN}$ is the maximum between $Z/Z_\odot$ and $0.01$. $Z$ is the metallicity of the injection region. Additionally, the gas within the injection region is fully ionized with gas temperature set as $10^4$ K. The kinetic and thermal energies of the gas are set to reflect this change. \\ 
\indent Along with the momentum, we also inject metals into the surrounding medium. The mass of the metals is calculated as the difference between the initial stellar mass and the helium core mass. The injected and subsequent mixing of these meals is modelled as a passive tracer field, where the metal tracer field follows the bulk gas flow. The metals are initially assigned to just the cells of the injection region, with each cell assigned a fractional proportion of metals according to its mass. The metallicity of a gas cell is defined as its metal mass divided by the gas mass. An important fact to note is that the metal tracer field is massless and doesn't interact with other particles through gravity.\\

\subsection{BH Formation\\}
After a SNe, the PopIII particle is converted to a BH particle depending on its ZAMS mass (For the purpose of this work, we ignore the BHs formed from PopII particles). 
\begin{enumerate}
    \item If the PopIII star has mass lower than 40 \msolar, the BH particle has the mass of the remnant helium core as discussed in the above section.
    \item For the direct collapse channel ($40< m_{\star}<140$ \msolar and $m_{\star}>260$ \msolar ), the entire PopIII particle mass is assigned to the BH particle.
    \item In the PISN channel ($140 < m_{\star} < 260$ \msolar), there is no BH remnant.
\end{enumerate}

\subsection{BH Accretion\\}
\label{sec:accretion}
\indent Once a PopIII particle has turned into a BH particle, we accrete onto it using the Bondi-Hoyle-Lyttleton (BHL) method \cite{bondi1952spherically, krumholz2004embedding}. Accretion onto a BH particle occurs by removing gas from the surrounding accretion sphere. The size of the accretion sphere is 5 cells wide. The accretion rate onto the BH particle is calculated using the usual Bondi formula
\begin{equation}
    \Dot{M}_{Bondi} = 4 \pi \rho_{\infty} r_{BH}^2 ((1.12 c_{\infty})^2 + v_{\infty}^2)^{1/2}
    \label{eq:bondi-hoyle-acc}
\end{equation}
where $\rho_{\infty}$ is the weighted density inside the accretion sphere, $r_{BH}=\frac{G M_{BH}}{v_{\infty}^2+c_{\infty}^2}$ is the BHL radius,  $v_{\infty}$ is the mass weighted velocity and $c_{\infty}$ is the sound speed of the surrounding gas. The term $v_{\infty}$ is calculated relative to the BH particle velocity and the sound speed is calculated only from the local region. The value of $\rho_{\infty}$ is calculated by assigning weights to all cells within the kernel radius \cite{krumholz2004embedding}, $r_K$, where the kernel radius is given by: 
\begin{equation}
\rm{r_K} = \left\{ \begin{array}{lcr}
  \Delta x/4 & &r_{BH} <  \Delta x/4\\
  r_{BH}  & & \ \ \ \ \Delta x/4 \le r_{BH} \le r_{acc}/2\\
  r_{acc}/2 && r_{BH} > r_{acc}/2
\end{array} \right.
\end{equation}
The weight assigned to the gas cells within the accretion sphere using is,
\begin{equation}
    \omega \propto \rm{exp(-r^2/r_{K}^2)}
\end{equation}
where $r$ is the distance from the cell from the BH particle. The proportionality symbol is used here to indicate that the actual weight assigned is relative to the other cells in the accretion sphere. The value of  $\rho_{\infty}$ is then assigned by applying the following formula:
\begin{equation}
     \rho_{\infty} = \bar{\rho} * \omega
\end{equation}
where $\bar{\rho}$ is the mass weighted mean density within the accretion sphere. This value of $\rho_{\infty}$ is used in calculating the accretion rate (Eqn \ref{eq:bondi-hoyle-acc}) at each timestep.  In order to improve our scheme further, we also include the impact of vorticity on our accretion rate.\\
\subsection{Vorticity\\}
The high-z universe has a chaotic nature, so we also account for the vorticity $\omega$ of the surrounding gas. The vorticity of the local medium is the mass-weighted average of the curl of gas cells with respect to the BH particle. From vorticity, we find a dimensionless vorticity $\omega_*$ given by,
\begin{equation}
\omega_* =  \omega \bigg( \frac{r_{BH}}{c_{\infty}} \bigg),
\end{equation}
which helps us calculate the damping factor $f(\omega)$ \cite{Krumholz_2006} as,
\begin{equation}
    f_{w} = \frac{1}{1 + \omega_*^{0.9}}
\end{equation}
and calculate the accretion rate in a turbulent medium according to,
\begin{equation}
    \Dot{M}_{\omega} = 4  \pi  \rho_{\infty}  r_{BH}^2  c_{\infty} * 0.34  f_{\omega_*}
\end{equation}
Finally, the total accretion rate onto the BH particle is,
\begin{equation}
    \Dot{M} = (\Dot{M}_{Bondi}^{-2} + \Dot{M}_{\omega}^{-2})^{-0.5}
\end{equation}
On a given time step, $t_h$, the BH particle accrete mass $t_h \times \Dot{M}$. For each gas cell, we calculate the fraction of mass lost to the BH particle. This fraction determines the linear momentum to be carried onto the BH particle such that the overall linear 
momentum is conserved. \\

\subsection{BH Accretion feedback\\}
\label{sec:thermal_feedback}
In nature, BHs return a fraction of accreted mass and energy to the surrounding gas in the form of both radiative and perhaps mechanical feedback via jets. In this study, we model feedback via the injection of thermal feedback which mimics the impact of radiative feedback on the surrounding material and to some (lesser) extend jet feedback (which may or may not be present in these LSBHs). The thermal energy is injected into the gas within the accretion region isotropically. The thermal energy is calculated based on the total mass accreted by the BH ($M_{acc}$) on that timestep by
\begin{equation} \label{eq:ThermalFeedback}
    E_{FB} = \epsilon_f f_c  \frac{M_{\rm acc}}{c^2},
\end{equation}
where $f_c$ is the thermal coupling factor set to 0.05 \cite{dimatteo2005energy, springel2005modelling, sijacki2007unified, dimatteo2008direct}, $\epsilon_f$ is the radiative efficiency for accretion and $\rm{M_{acc}}$ is the accreted mass on this timestep. For sub-Eddington accretion, we calculate $\epsilon_f$ as \cite{sadowski2016energy}
\begin{equation}
    \epsilon_f = 1 - \sqrt{1-\frac{2}{3 R_{\rm ISCO}}},
\label{eq:epsilon}
\end{equation}
where $R_{\rm ISCO}$ is the innermost stable orbit, which is dependent on the spin, $a$ of
the black hole. We assign $a = 0.7$ \cite{abramowicz2013foundations} in all cases which gives us $\epsilon = 0.1$. \\
\indent The calculation of $\epsilon$ changes for super-Eddington accretion rates.
In this case, we follow the prescription given in Regan et al. (2019) \cite{Regan_2019} which itself uses the parametrisation given in Madau, Haardt \& Dotti (2014)\cite{Madau_2014}.
The super-Eddington luminosity is then given by
\begin{equation}
    L_{\rm SE}=L_{\rm Edd} A \left(  \frac{0.985}{r_{\rm Edd} + B} + \frac{0.015}{r_{\rm Edd} + C} \right),
\end{equation}
where $r_{\rm Edd}$ is the ratio of of the accretion rate to the Eddington limit, and the variables $A, B, C$ depend on the black hole spin parameter $a$. 
\begin{equation}
\begin{aligned}
A =& (0.9663 - 0.9292  a)^{-0.5639}, \\
B =& (4.627 - 4.445  a)^{-0.5524},\\
C =& (827.3 - 718.1  a)^{-0.7060},\\
\end{aligned}
\end{equation}
As before we set $a$ is 0.7. $L_{SE}$ is the super-Eddington luminosity value with $L_{Edd}$ being the canonical Eddington luminosity with
\begin{equation}
    L_{\rm Edd}=\frac{4 \pi G M m_p \mu c}{\sigma_{\rm T}},
\end{equation}

Finally we calculate the super-Eddington radiative efficiency $\epsilon$ as
\begin{equation}
\epsilon  = L_{\rm SE} / (\dot{M} c^2).
\end{equation}

and use that in Equation \ref{eq:ThermalFeedback} for the super-Eddington limit.


\subsection{BH mergers\\}
Alongside accretion, BHs can also grow through mergers with other BHs. In our work, we allow for BHs to merge if they satisfy the following conditions \cite{prole2022fragmentation}:
\begin{enumerate}
    \item They lie within each other’s accretion radius.
    \item They are moving towards each other.
    \item Their relative accelerations are $<0$.
    \item They are gravitationally bound to each other.
\end{enumerate}

If the above criterion are satisfied, we merge the two BHs, conserving mass and linear momentum. Post-merger, the heavier BHs gains the physical properties of the combined remnant while the secondary BH is deleted from the simulation.

\subsection{Chemistry solver\\}
We model the chemistry using SGCHEM \cite{hartwig2015improved, wollenberg2020formation, prole2022fragmentation}, which consists of 12 chemical species, H, H\textsuperscript{+}, H\textsuperscript{-}, H\textsubscript{2}\textsuperscript{+} , H\textsubscript{2} , He, He\textsuperscript{+}, He\textsuperscript{++}, D, D\textsuperscript{+}, HD, and e\textsuperscript{-}. Based on the chemical network described in Clark et al. 2011 \cite{clark2011formation} and updated in Schauer et al. 2017 \cite{schauer2017formation}, the primordial gas follows 45 reactions. We set the initial gas fractions at $2\times10^{-6}$ for H\textsubscript{2}, $10^{-4}$ for H\textsuperscript{+}, $2.6\times10^{-9}$ for D\textsuperscript{+}, and $2.6\times10^{-5}$ for D with respect to neutral hydrogen. The chemical network accounts for H\textsubscript{2} cooling, HD cooling, heating and cooling from gas, shocks, compression, and expansion of the gas, collisionally induced H\textsubscript{2} emission, ionisation, and recombination.

\begin{figure}
    \centering
    \includegraphics[width=\linewidth]{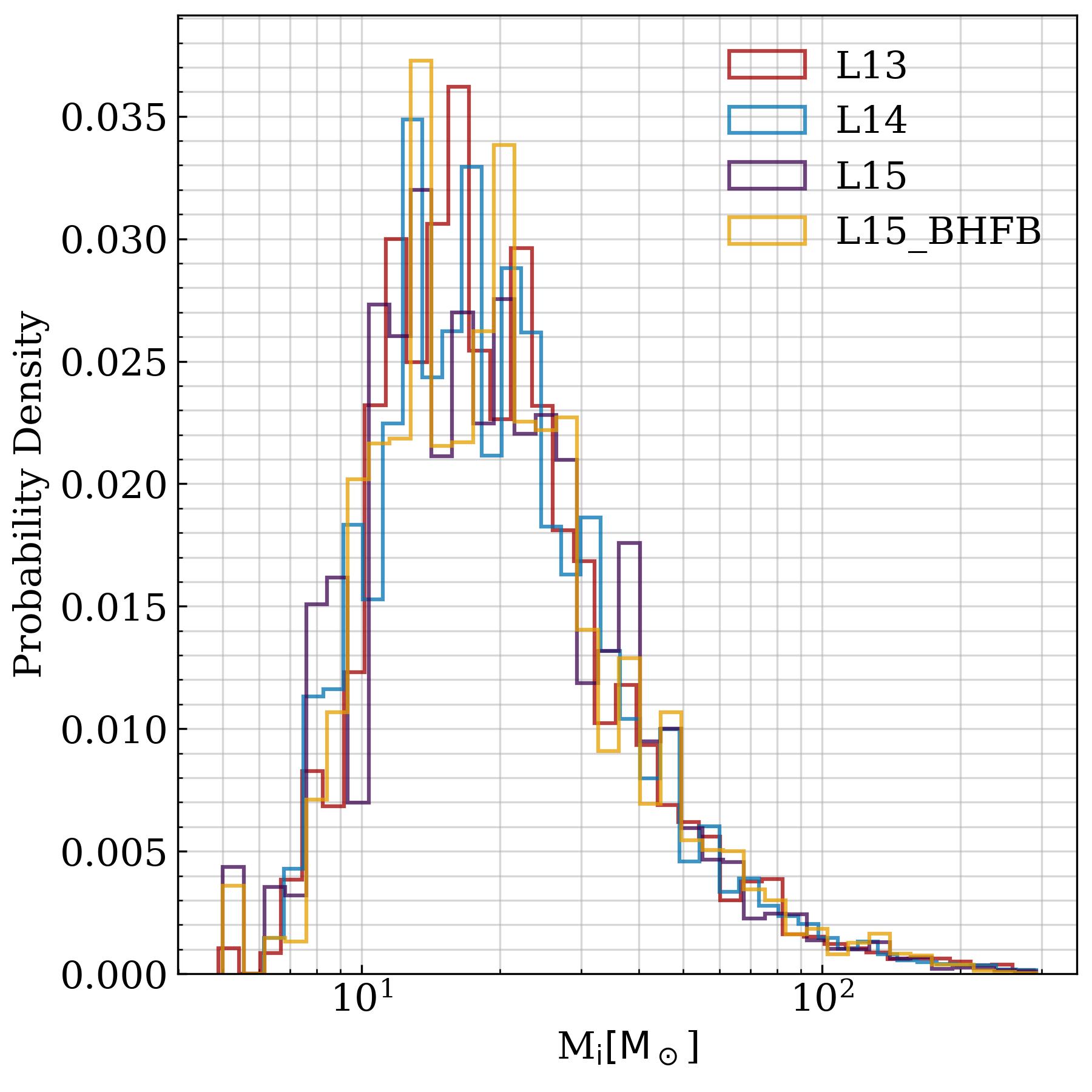}
    \caption{\textbf{Extended Data Figure 1 | Population density of PopIII stars.} We show the IMF of PopIII stars from all our simulations. The simulations L13, L14, L15 and L15\_BHFB are coloured red, blue, violet and yellow respectively. This is the top-heavy IMF with a characteristic mass of 20 \msolarc. The minimum mass of PopIII stars in 1 \msolar and the maximum mass is 300 \msolarc.}
    \label{fig:Mehta_ED_Fig1}
\end{figure}

\begin{figure}
    \centering
    \includegraphics[width=\linewidth]{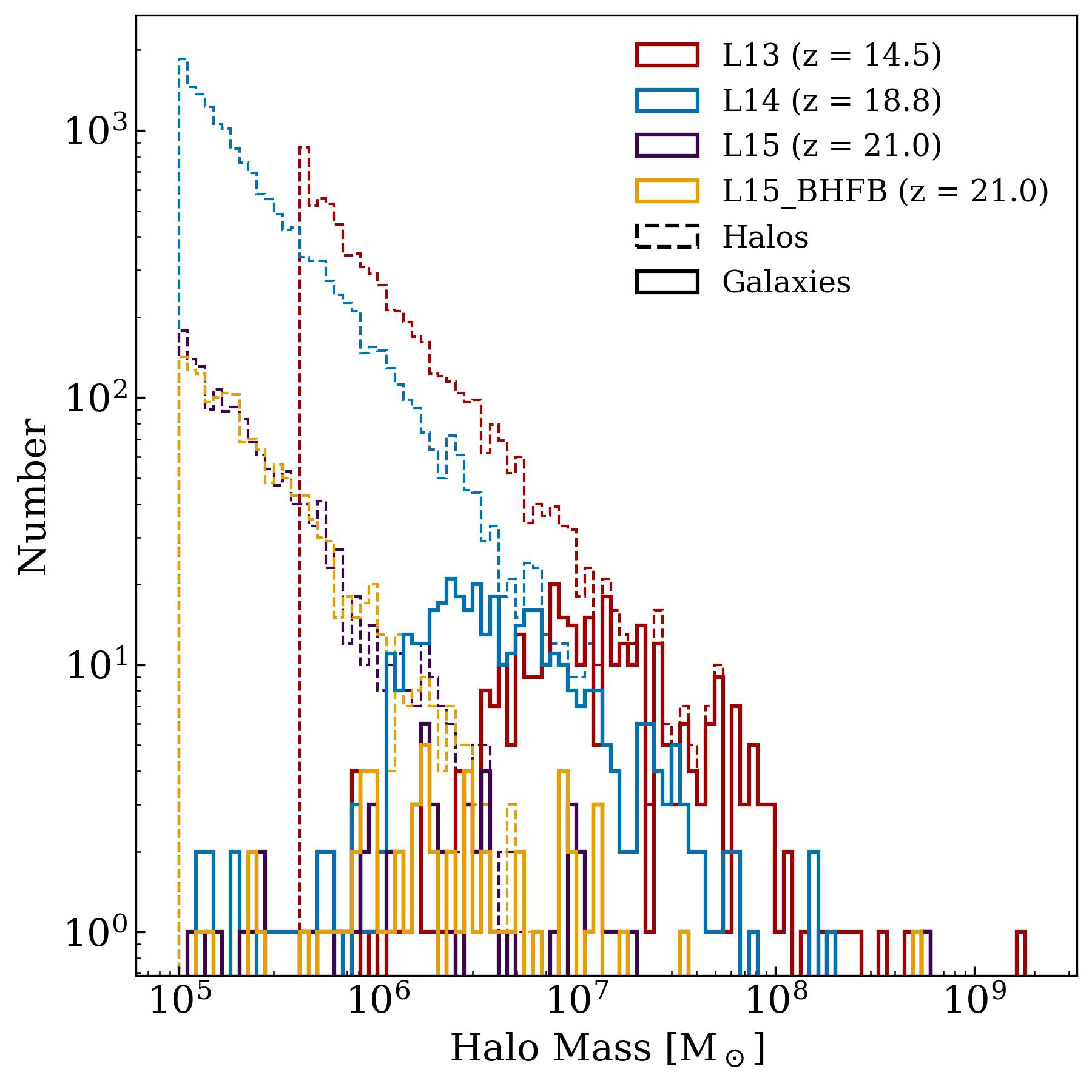}
    \caption{\textbf{Extended Data Figure 2 |Mass function of halos and galaxies.} We show the number of halos (dashed lines) identified through the FOF algorithm across each of our simulations. The simulations L13, L14, L15 and L15\_BHFB are coloured red, blue, violet and yellow respectively. We also highlight the number of galaxies (solid lines) within the halos that went on to have star formation and host BHs. We do not count here halos with mass $< 10^5$ \msolar. We find that with increasing resolution, star formation begins in smaller and smaller galaxies.}
    \label{fig:Mehta_ED_Fig2}
\end{figure}

\begin{figure}
    \centering
    \includegraphics[width=\linewidth]{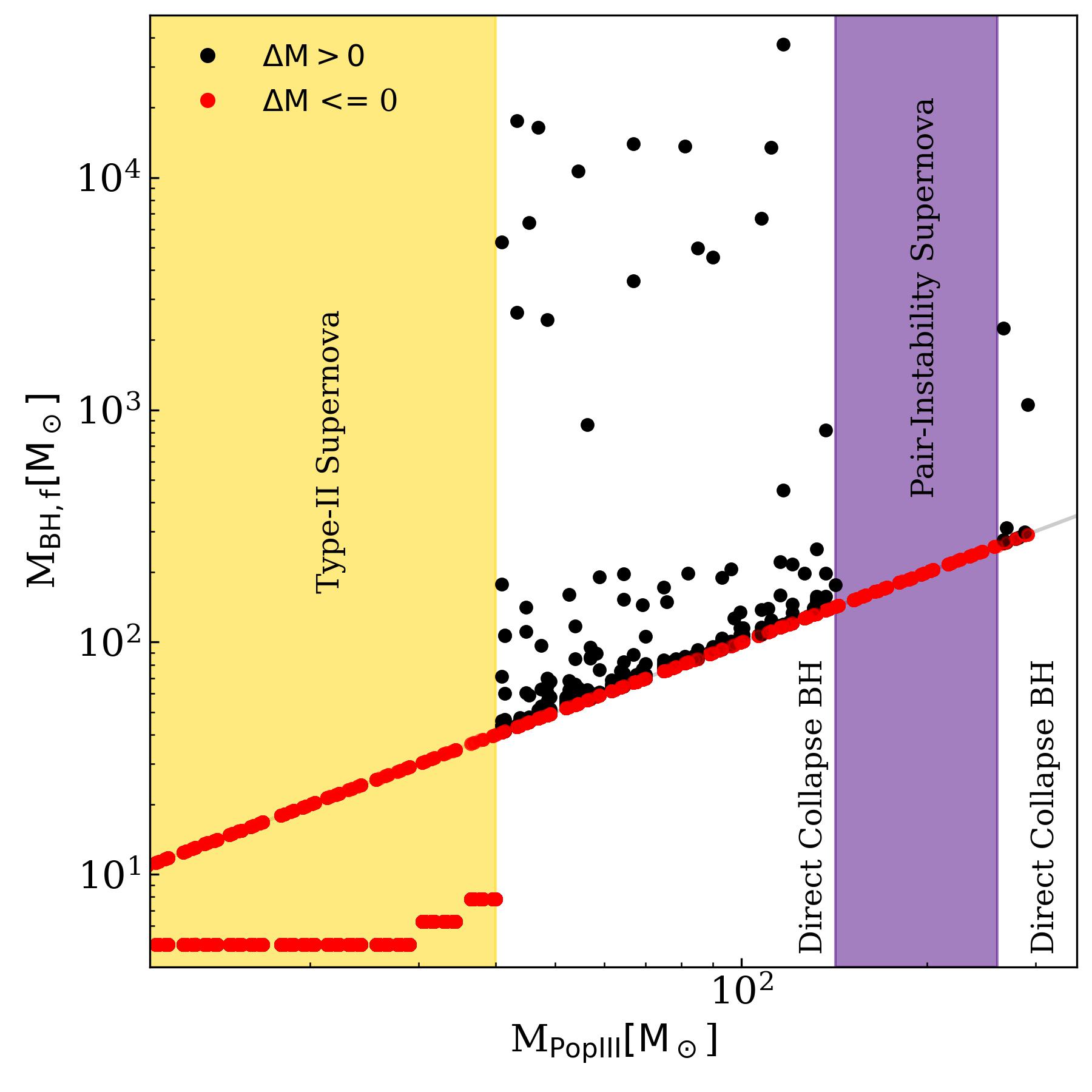}
    \caption{\textbf{Extended Data Figure 3 |Final mass of BHs vs initial mass of their PopIII progenitor stars.} We show the final masses of BHs as a function of their progenitor PopIII stars. All black dots are BHs that grew larger than their progenitor mass while red dots are BHs that did not accrete any gas. The shaded patch in yellow denotes the BH formed after undergoing a Type-II supernova. The shaded purple patch is for PISN, so there are no BH remnants. While in the white patch are BHs formed through the direct collapse channel. Interestingly, all of the BHs that grow formed through the direct collapse channel.}
    \label{fig:Mehta_ED_Fig3}
\end{figure}

\begin{figure*}
    \centering
    \includegraphics[width=\linewidth, height=16cm]{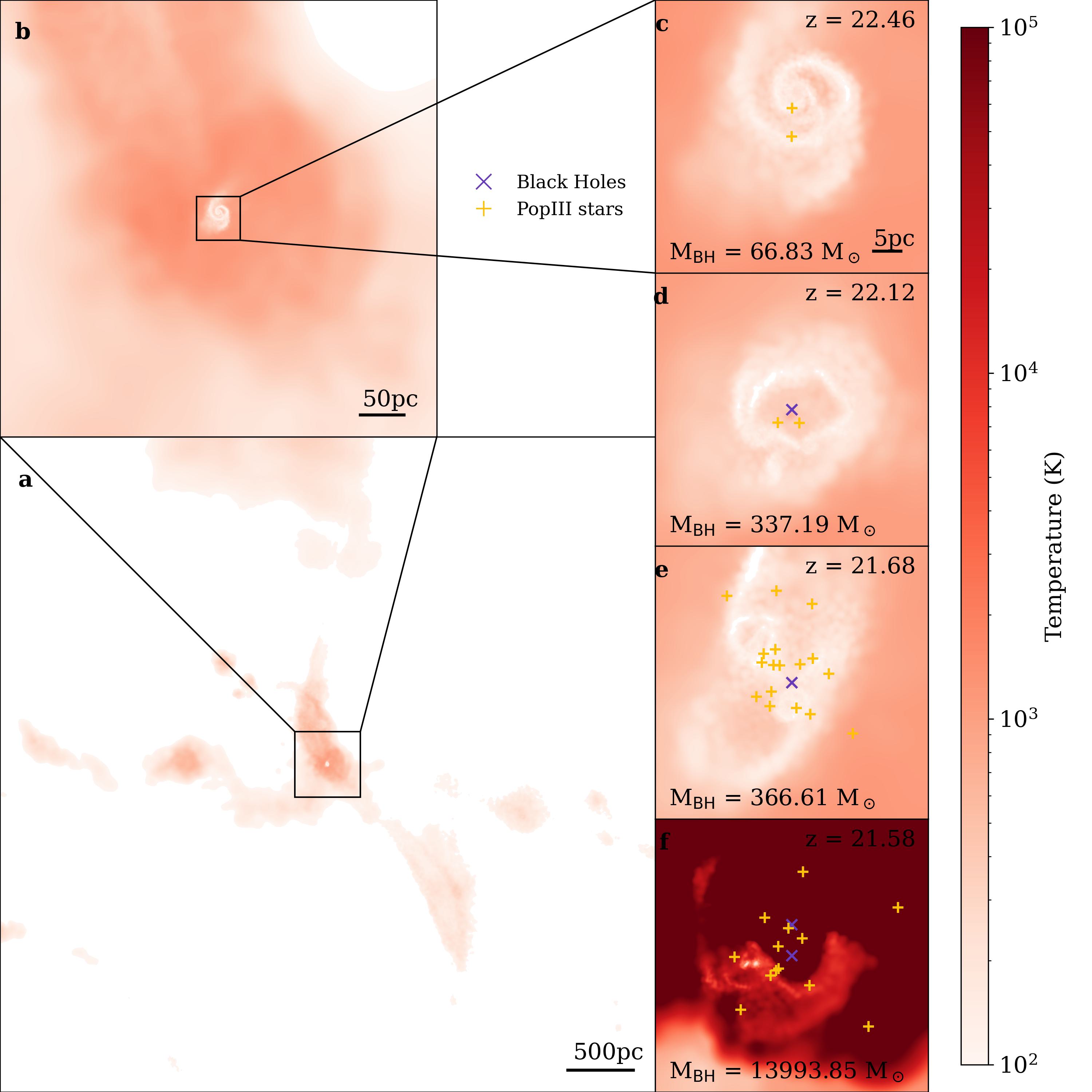}
    \caption{\textbf{Extended Data Figure 4 |Gas temperature projection for PopIII star formation, supernova feedback, and BH feedback for the most massive BH in the L15\_BHFB simulations.} a, b, c: Gas collapses to form cold galaxies in hot cosmic filaments. Within these galaxies, PopIII star formation occurs. d: The first PopIII star transitions to a BH and accretes the surrounding mass, simultaneously injecting thermal energy back into the gas. This thermal energy causes the galaxy to be distorted. e: The galaxy eventually reassembles triggering a second epoch of star formation. f: SNe from PopIII stars and BH thermal feedback heats up the gas to $10^5$ K, which halts further BH growth and star formation.}
    \label{fig:Mehta_ED_Fig4}
\end{figure*}

\begin{figure}
    \centering
    \includegraphics[width=\linewidth]{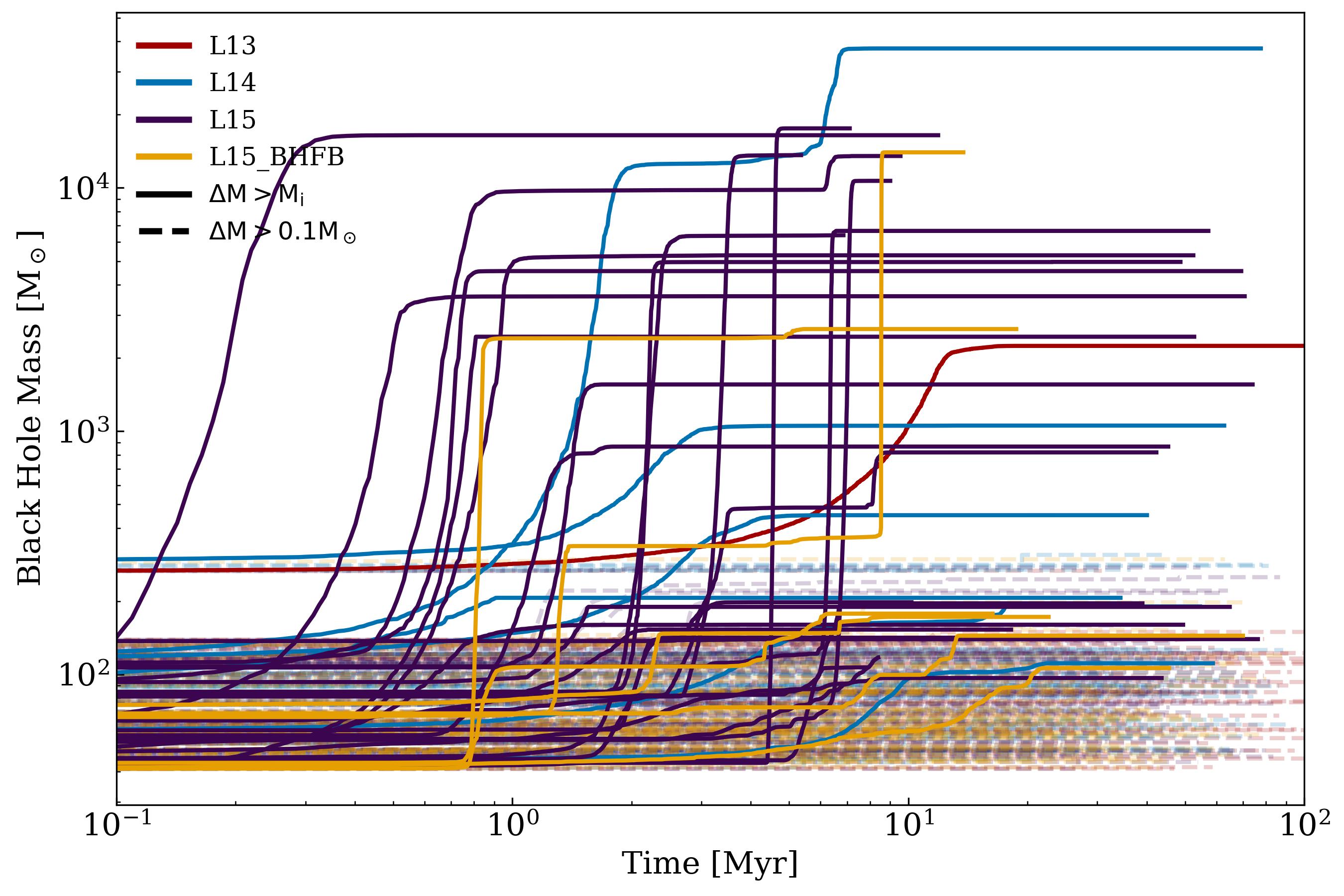}
    \caption{\textbf{Extended Data Figure 5 |BH particle growth plots.} We show the growth of BH particles with time after they transition from PopIII particles for BHs that accreted more than 0.1 \msolar(dashed) and BHs that doubled their initial mass (solid lines). The simulations L13, L14, L15 and L15\_BHFB are coloured red, blue, violet and yellow respectively. On average, we the find the growth phase lasting around a million years. We also see a trend that with increasing resolution, the timescales shorten.}
    \label{fig:Mehta_ED_Fig5}
\end{figure}

\begin{figure}
    \centering
    \includegraphics[width=\linewidth]{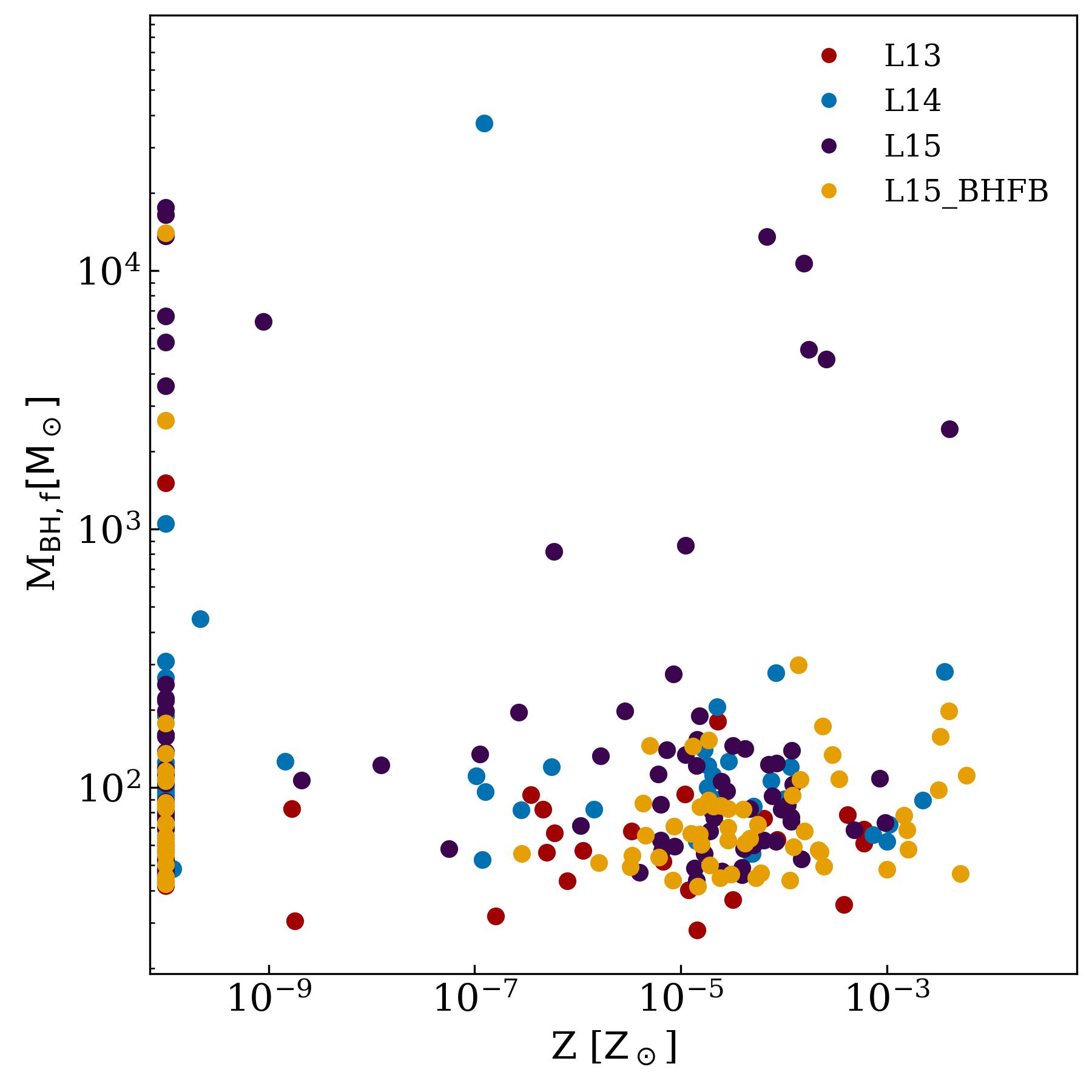}
    \caption{\textbf{Extended Data Figure 6 |Average metallicity of gas surrounding the BHs.} We show the average metallicity of gas surrounding the BHs during its entire growth phase. The simulations L13, L14, L15 and L15\_BHFB are coloured red, blue, violet and yellow respectively. In our simulations, we have a metallicity floor of $10^{-10}$ $\rm{Z}_\odot$. We have a subset of BHs growing in pristine metal-free conditions, but we also see growth of BHs in galaxies enriched with metals.}
    \label{fig:Mehta_ED_Fig6}
\end{figure}

\begin{figure*}
    \centering
    \includegraphics[width=\linewidth]{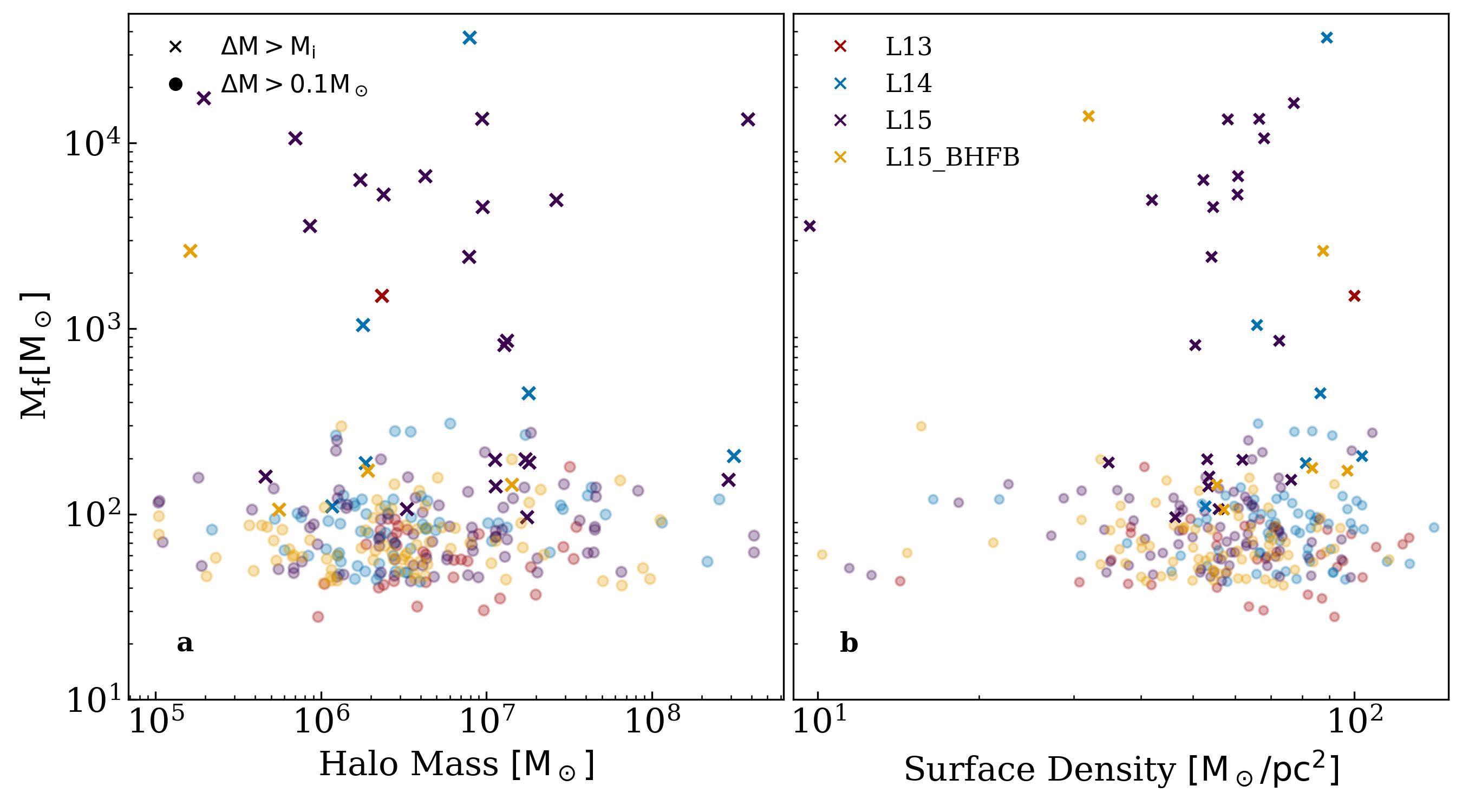}
    \caption{\textbf{Extended Data Figure 7 |Relation between BH growth and halo properties.} a: We show the final masses of BHs (M$_{\mathrm{f}}$) as a function of its host halo mass. The simulations L13, L14, L15 and L15\_BHFB are coloured red, blue, violet and yellow respectively. The dots show BHs that accreted more than 0.1 \msolar and crosses focus on BHs that accreted more than their initial mass (M$_{\mathrm{i}}$). We see no correlation between the two, suggesting that the halo does not need to be special in order for BHs to grow. b: We see a similar result when comparing the final masses with surface density of gas surrounding the BHs. However, we are incomplete in sampling larger halo masses and it is possible, even likely, that more massive halos would support additional growth and will also likely support the growth of LSBHs which have already experienced previous growth episodes.}
    \label{fig:Mehta_ED_Fig7}
\end{figure*}

\begin{figure}
    \centering
    \includegraphics[width=\linewidth]{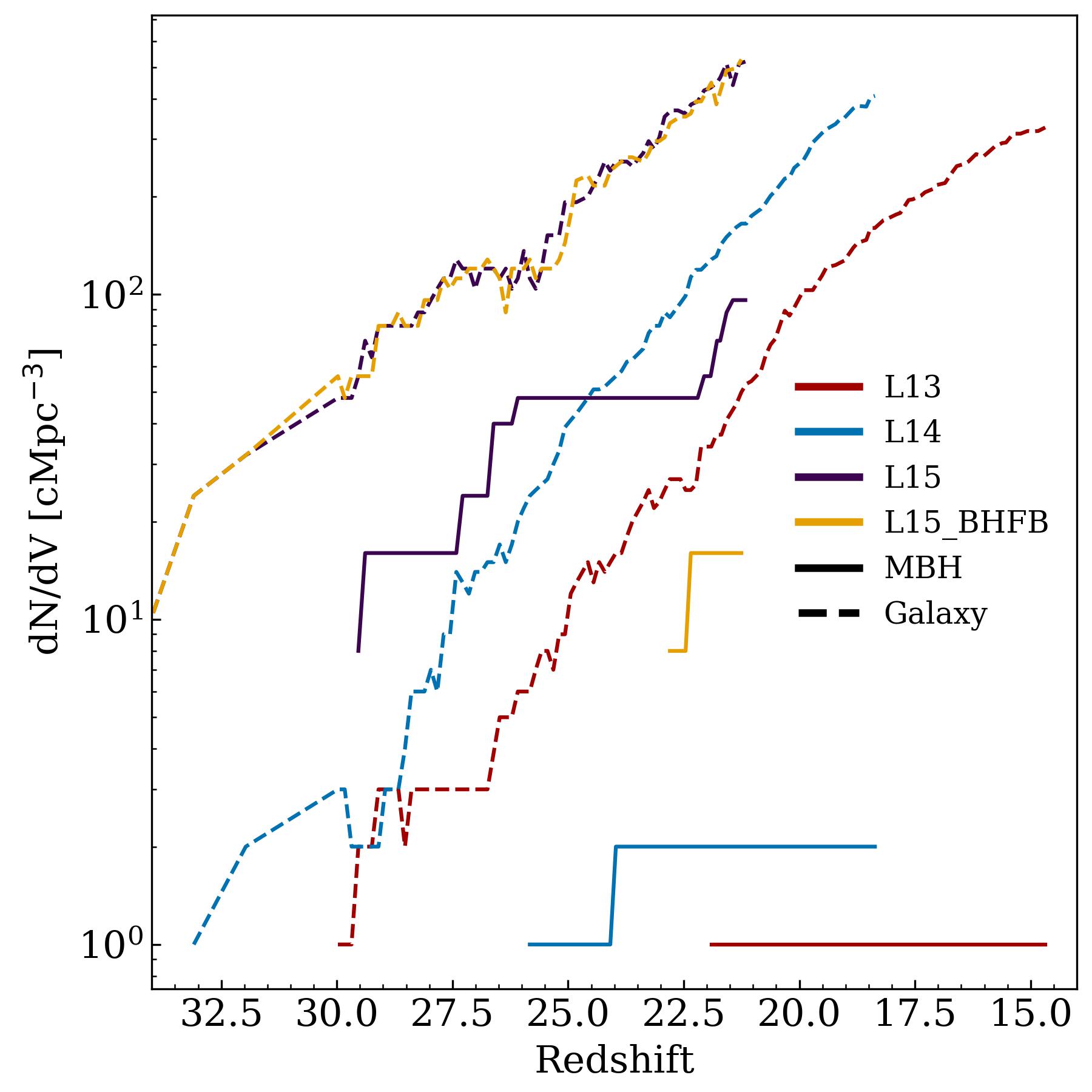}
    \caption{\textbf{Extended Data Figure 8 |Number density of MBHs and galaxies in our simulations.} We show the number density evolution of BHs above 1000 \msolar (solid lines) along with the galaxy (dashed lines) number density. The simulations L13, L14, L15 and L15\_BHFB are coloured red, blue, violet and yellow respectively. The plot highlights one of our results, that increasing resolution helps us capture LSBH growth. The number density of BHs reaches almost 100 $\rm{cMpc^{-3}}$ for the L15 simulation. In the L15\_BHFB simulation, the number density decreases to below 20 $\rm{cMpc^{-3}}$, which is still much larger than the number density of the high-z AGN populations $\lesssim 10^{-2}$ $\rm{cMpc^{-3}}$, making LSBHs promising progenitors for SMBHs.}
    \label{fig:Mehta_ED_Fig8}
\end{figure}
\end{methods}


\subsection{References for Methods}
\bibliographystyle{naturemag}

\begin{thebibliography}{99}

\bibitem{madau2001massive}
Madau, P., \& Rees, M.~J.
Massive Black Holes as Population III Remnants.
\textit{\apjl}, \textbf{551(1)}, L27–L30 (2001).
doi:10.1086/319848.

\bibitem{volonteri2003assembly}
Volonteri, M., Haardt, F., \& Madau, P.
The Assembly and Merging History of Supermassive Black Holes in Hierarchical Models of Galaxy Formation.
\textit{\apj}, \textbf{582(2)}, 559–573 (2003).
doi:10.1086/344675.

\bibitem{madau2004early}
Madau, P., Rees, M.~J., Volonteri, M., Haardt, F., \& Oh, S.~P. 
Early Reionization by Miniquasars.
\textit{\apj}, \textbf{604(2)}, 484–494 (2004).
doi:10.1086/381935.

\bibitem{Volonteri_2012}
Volonteri, M.
The Formation and Evolution of Massive Black Holes.
\textit{\sci}, \textbf{337(6094)}, 544 (2012).
doi:10.1126/science.1220843.

\bibitem{Latif_2022}
Latif, M.~A., Whalen, D., \& Khochfar, S.
The Birth Mass Function of Population III Stars.
\textit{\apj}, \textbf{925(1)}, 28 (2022).
doi:10.3847/1538-4357/ac3916.

\bibitem{abel2002formation}
Abel, T., Bryan, G.~L., \& Norman, M.~L. 
The Formation of the First Star in the Universe.
\textit{\sci}, \textbf{295(5552)}, 93–98 (2002).
doi:10.1126/science.1063991.

\bibitem{bromm2002formation}
Bromm, V., Coppi, P.~S., \& Larson, R.~B.
The Formation of the First Stars. I. The Primordial Star-forming Cloud.
\textit{\apj}, \textbf{564(1)}, 23–51 (2002).
doi:10.1086/323947.

\bibitem{o2007population}
O'Shea, B.~W., \& Norman, M.~L. 
Population III Star Formation in a $\Lambda$CDM Universe. II. Effects of a Photodissociating Background.
\textit{\apj}, \textbf{673(1)}, 14–33 (2008).
doi:10.1086/524006.

\bibitem{turk2009formation}
Turk, M.~J., Abel, T., \& O'Shea, B. 
The Formation of Population III Binaries from Cosmological Initial Conditions.
\textit{\sci}, \textbf{325(5940)}, 601 (2009).
doi:10.1126/science.1173540.

\bibitem{hirano2014one}
Hirano, S., Hosokawa, T., Yoshida, N., Umeda, H., Omukai, K., Chiaki, G., \& Yorke, H.~W. 
One Hundred First Stars: Protostellar Evolution and the Final Masses.
\textit{\apj}, \textbf{781(2)}, 60 (2014).
doi:10.1088/0004-637X/781/2/60.

\bibitem{prole2023dark}
Prole, L.~R., Schauer, A.~T.~P., Clark, P.~C., Glover, S.~C.~O., Priestley, F.~D., \& Klessen, R.~S. 
From dark matter halos to pre-stellar cores: high resolution follow-up of cosmological Lyman-Werner simulations.
\textit{\mnras}, \textbf{520(2)}, 2081–2093 (2023).
doi:10.1093/mnras/stad188.

\bibitem{woosley1995evolution}
Woosley, S.~E., \& Weaver, T.~A. 
The Evolution and Explosion of Massive Stars. II. Explosive Hydrodynamics and Nucleosynthesis.
\textit{\apjs}, \textbf{101}, 181 (1995).
doi:10.1086/192237.

\bibitem{nomoto2006nucleosynthesis}
Nomoto, K., Tominaga, N., Umeda, H., Kobayashi, C., \& Maeda, K. 
Nucleosynthesis yields of core-collapse supernovae and hypernovae, and galactic chemical evolution.
\textit{Nucl. Phys. A}, \textbf{777}, 424–458 (2006).
doi:10.1016/j.nuclphysa.2006.05.008.

\bibitem{heger2002nucleosynthetic}
Heger, A., \& Woosley, S.~E.
The Nucleosynthetic Signature of Population III.
\textit{\apj}, \textbf{567(1)}, 532–543 (2002).
doi:10.1086/338487.

\bibitem{Lupi_2016}
Lupi, A., Haardt, F., Dotti, M., Fiacconi, D., Mayer, L., \& Madau, P.
Growing massive black holes through supercritical accretion of stellar-mass seeds.
\textit{\mnras}, \textbf{456(3)}, 2993–3003 (2016).
doi:10.1093/mnras/stv2877.

\bibitem{smith2018growth}
Smith, B.~D., Regan, J.~A., Downes, T.~P., Norman, M.~L., O'Shea, B.~W., \& Wise, J.~H. 
The growth of black holes from Population III remnants in the Renaissance simulations.
\textit{\mnras}, \textbf{480(3)}, 3762–3773 (2018).
doi:10.1093/mnras/sty2103.

\bibitem{Regan_2019}
Regan, J.~A., Downes, T.~P., Volonteri, M., Beckmann, R., Lupi, A., Trebitsch, M., \& Dubois, Y. 
Super-Eddington accretion and feedback from the first massive seed black holes.
\textit{\mnras}, \textbf{486(3)}, 3892–3906 (2019).
doi:10.1093/mnras/stz1045.

\bibitem{sassano2023super}
Sassano, F., Capelo, P.~R., Mayer, L., Schneider, R., \& Valiante, R.
Super-critical accretion of medium-weight seed black holes in gaseous proto-galactic nuclei.
\textit{\mnras}, \textbf{519(2)}, 1837–1855 (2023).
doi:10.1093/mnras/stac3608.

\bibitem{inayoshi2016hyper}
Inayoshi, K., Haiman, Z., \& Ostriker, J.~P.
Hyper-Eddington accretion flows on to massive black holes.
\textit{\mnras}, \textbf{459(4)}, 3738–3755 (2016).
doi:10.1093/mnras/stw836.

\bibitem{jiang2019super}
Jiang, Y.-F., Stone, J.~M., \& Davis, S.~W.
Super-Eddington Accretion Disks around Supermassive Black Holes.
\textit{\apj}, \textbf{880(2)}, 67 (2019).
doi:10.3847/1538-4357/ab29ff.

\bibitem{park2020biconical}
Park, K., Wise, J. H., Bogdanovic, T., \& Ricotti, M.
Biconical-dominated Accretion Flow onto Seed Black Holes in a Hyperaccretion Regime.
\textit{\apj}, \textbf{905}, 92 (2020). 
doi:10.3847/1538-4357/abc336.

\bibitem{kitaki2021origins}
Kitaki, T., Mineshige, S., Ohsuga, K., \& Kawashima, T.
The origins and impact of outflow from super-Eddington flow.
\textit{\pasj}, \textbf{73}, 450–466 (2021). 
doi:10.1093/pasj/psab011.

\bibitem{botella2022structure}
Botella, I., Mineshige, S., Kitaki, T., Ohsuga, K., \& Kawashima, T.
Structure of the super-Eddington outflow and its impact on the cosmological scale.
\textit{\pasj}, \textbf{74}, 384–397 (2022). 
doi:10.1093/pasj/psac001.

\bibitem{lambrides2024case}
Lambrides, E., Garofali, K., Larson, R., et al.
The Case for Super-Eddington Accretion: Connecting Weak X-ray and UV Line Emission in JWST Broad-Line AGN During the First Gyr of Cosmic Time.
\textit{arXiv e-prints (2024)}. 
doi:10.48550/arXiv.2409.13047.

\bibitem{suh2025super}
Suh, H., Scharwachter, J., Farina, E. P., et al.
A super-Eddington-accreting black hole $\sim$1.5 Gyr after the Big Bang observed with JWST.
\textit{\nastro}, \textbf{9}, 271–279 (2025). 
doi:10.1038/s41550-024-02402-9.

\bibitem{shi2023hyper}
Shi, Y., Kremer, K., Grudic, M. Y., Gerling-Dunsmore, H. J., \& Hopkins, P. F.
Hyper-Eddington black hole growth in star-forming molecular clouds and galactic nuclei: can it happen?
\textit{\mnras}, \textbf{518}, 3606–3621 (2023). 
doi:10.1093/mnras/stac3245.

\bibitem{gordon2024hungry}
Gordon, S. T., Smith, B. D., Khochfar, S., \& Regan, J. A.
Hungry or not: how stellar-mass black holes grow (or don't) in dark matter mini-haloes at high resolution.
\textit{\mnras}, \textbf{529}, 604–627 (2024). 
doi:10.1093/mnras/stae566.

\bibitem{mehta2024growth}
Mehta, D., Regan, J. A., \& Prole, L.
Growth of Light-Seed Black Holes in Gas-Rich Galaxies at High Redshift.
\textit{The Open Journal of Astrophysics}, \textbf{7}, 107 (2024). doi:10.33232/001c.126629.

\bibitem{Alexander_2014}
Alexander, T., \& Natarajan, P.
Rapid growth of seed black holes in the early universe by supra-exponential accretion.
\textit{\sci} \textbf{345}, 1330-1333 (2014). 
doi:10.1126/science.1251053.

\bibitem{zana2025super}
Zana, T., Capelo, P. R., Boresta, M., Schneider, R., Lupi, A., Trinca, A., Mayer, L., Valiante, R., \& Graziani, L.
Super-Eddington accretion in protogalactic cores.
\textit{arXiv e-prints (2025)}. 
doi:10.48550/arXiv.2508.21114.

\bibitem{Alvarez_2009}
Alvarez, M.A., Wise, J.~H., \& Abel, T.
Accretion onto the First Stellar-Mass Black Holes.
\textit{\apjl} \textbf{701}, L133-L137 (2009).
doi:10.1088/0004-637X/701/2/L133.

\bibitem{Springel_2010}
Springel, V.
E pur si muove: Galilean-invariant cosmological hydrodynamical simulations on a moving mesh.
\textit{\mnras}, \textbf{401}, 791–851 (2010). 
doi:10.1111/j.1365-2966.2009.15715.x.

\bibitem{Pakmor_2016}
Pakmor, Ruediger, Springel, V., Bauer, A., Mocz, P., Munoz, D. J., Ohlmann, S. T., Schaal, K., \& Zhu, C.
Improving the convergence properties of the moving-mesh code AREPO.
\textit{\mnras}, \textbf{455}, 1134–1143 (2016). 
doi:10.1093/mnras/stv2380.

\bibitem{hahn2011multi}
Hahn, O., \& Abel, T.
Multi-scale initial conditions for cosmological simulations.
\textit{\mnras}, \textbf{415}, 2101–2121 (2011).
doi:10.1111/j.1365-2966.2011.18820.x

\bibitem{Prole_2025}
Prole, L. R., Regan, J. A., Mehta, D., Coles, P., \& Dayal, P.
Primordial black holes in cosmological simulations: growth prospects for supermassive black holes.
\textit{arXiv e-prints (2025)}. 
doi:10.48550/arXiv.2506.11233.

\bibitem{Prole_2025b}
Prole, L. R., Regan, J. A., Mehta, D., Pakmor, R., Koudmani, S., Bourne, M. A., Glover, S. C. O., Wise, J. H., Klessen, R. S., Tremmel, M., Sijacki, D., Beckmann, R. S., Haehnelt, M. G., Brennan, J., van de Bor, P., \& Clark, P. C.
The SEEDZ Simulations: Methodology and First Results on Massive Black Hole Seeding and Early Galaxy Growth.
\textit{arXiv e-prints (2025)}. 
doi:10.48550/arXiv.2511.09640.

\bibitem{qin2020tale}
Qin, Y., Mesinger, A., Park, J., Greig, B., \& Munoz, J. B.
A tale of two sites–I. Inferring the properties of minihalo-hosted galaxies from current observations.
\textit{\mnras}, \textbf{495}, 123–140 (2020).
doi:10.1093/mnras/staa1131

\bibitem{gordon2025conditions}
Gordon, S. T., Smith, B. D., Khochfar, S., \& Beckmann, R. S.
Conditions for super-Eddington accretion onto the first black holes.
\textit{\mnras}, \textbf{537}, 674–690 (2025). 
doi:10.1093/mnras/staf054.

\bibitem{Regan_2020b}
Regan, J.~A., Wise, J.~H., Woods, T.~E., Downes, T.~P., O'Shea, B.~W., \& Norman, M.~L.
(2020). The Formation of Very Massive Stars in Early Galaxies and Implications for Intermediate Mass Black Holes.
\textit{The Open Journal of Astrophysics}, \textbf{3(1)}, 15 (2020). doi:10.21105/astro.2008.08090

\bibitem{jaura2022trapping}
Jaura, O., Glover, S.~C.~O., Wollenberg, K.~M.~J., Klessen, R.~S., Geen, S., \& Haemmerlé, L.
(2022). Trapping of H II regions in Population III star formation.
\textit{\mnras}, \textbf{512(1)}, 116--136 (2022). 
doi:10.1093/mnras/stac487

\bibitem{volonteri2015case}
Volonteri, M., Silk, J., \& Dubus, G.
(2015). The Case for Supercritical Accretion onto Massive Black Holes at High Redshift.
\textit{\apj}, \textbf{804(2)}, 148 (2015). 
doi:10.1088/0004-637X/804/2/148

\bibitem{jiang2014global}
Jiang, Y.-F., Stone, J.~M., \& Davis, S.~W.
(2014). A Global Three-dimensional Radiation Magneto-hydrodynamic Simulation of Super-Eddington Accretion Disks.
\textit{\apj}, \textbf{796(2)}, 106 (2014). 
doi:10.1088/0004-637X/796/2/106

\bibitem{Tremmel_2015}
Tremmel, M., Governato, F., Volonteri, M., \& Quinn, T.~R.
(2015). Off the beaten path: a new approach to realistically model the orbital decay of supermassive black holes in galaxy formation simulations.
\textit{\mnras}, \textbf{451}, 1868--1874 (2015). 
doi:10.1093/mnras/stv1060

\bibitem{Pfister_2019}
Pfister, H., Volonteri, M., Dubois, Y., Dotti, M., \& Colpi, M.
(2019). The erratic dynamical life of black hole seeds in high-redshift galaxies.
\textit{\mnras}, \textbf{486(1)}, 101--111 (2019). 
doi:10.1093/mnras/stz822

\bibitem{habouzit2017blossoms}
Habouzit, M., Volonteri, M., \& Dubois, Y.
(2017). Blossoms from black hole seeds: properties and early growth regulated by supernova feedback.
\textit{\mnras}, \textbf{468(4)}, 3935--3948 (2017). 
doi:10.1093/mnras/stx666

\bibitem{bhowmick2024introducing}
Bhowmick, A.~K., Blecha, L., Torrey, P., Kelley, L.~Z., Weinberger, R., Vogelsberger, M., Hernquist, L., Somerville, R.~S., \& Evans, A.~E.
(2024). Introducing the BRAHMA simulation suite: signatures of low-mass black hole seeding models in cosmological simulations.
\textit{\mnras}, \textbf{531(4)}, 4311--4335 (2024). 
doi:10.1093/mnras/stae1386

\bibitem{taylor2014seeding}
Taylor, P., \& Kobayashi, C.
(2014). Seeding black holes in cosmological simulations.
\textit{\mnras}, \textbf{442(3)}, 2751--2767 (2014). 
doi:10.1093/mnras/stu983

\bibitem{perez2024nature}
P\'erez-Gonz\'alez, P.~G., Barro, G., Rieke, G.~H., Lyu, J., Rieke, M., Alberts, S., Williams, C.~C., Hainline, K., Sun, F., Pusk\'as, D., Annunziatella, M., Baker, M.~W., Bunker, A.~J., Egami, E., Ji, Z., Johnson, B.~D., Robertson, B., Rodr\'iguez Del Pino, B., Rujopakarn, W., Shivaei, I., Tacchella, S., Willmer, C.~N.~A., \& Willott, C.
(2024). What Is the Nature of Little Red Dots and what Is Not, MIRI SMILES Edition.
\textit{\apj}, \textbf{968(1)}, 4 (2024). 
doi:10.3847/1538-4357/ad38bb

\bibitem{kokorev2024census}
Kokorev, V., Caputi, K.~I., Greene, J.~E., Dayal, P., Trebitsch, M., Cutler, S.~E., Fujimoto, S., Labb\'e, I., Miller, T.~B., Iani, E., Navarro-Carrera, R., \& Rinaldi, P.
(2024). A Census of Photometrically Selected Little Red Dots at 4 < z < 9 in JWST Blank Fields.
\textit{\apj}, \textbf{968(1)}, 38 (2024). 
doi:10.3847/1538-4357/ad4265

\bibitem{shi2024feedback}
Shi, Y., Kremer, K., \& Hopkins, P. F.
Feedback-regulated seed black hole growth in star-forming molecular clouds and galactic nuclei.
\textit{\aap}, \textbf{691}, A24 (2024). doi:10.1051/0004-6361/202450964.

\end{thebibliography}

\begin{thebibliography}{99}

\bibitem{Springel_2021}
Springel, V., Pakmor, R., Zier, O. \& Reinecke, M.  
Simulating cosmic structure formation with the GADGET-4 code.  
\textit{\mnras} \textbf{506}, 2871–2949 (2021).
doi:10.1093/mnras/stab1855

\bibitem{bourne2024dynamics}
Bourne, M. A., Fiacconi, D., Sijacki, D., Piotrowska, J. M. \& Koudmani, S.  
Dynamics and spin alignment in massive, gravito-turbulent circumbinary discs around supermassive black hole binaries.  
\textit{\mnras} \textbf{534}, 3448–3477 (2024).
doi:10.1093/mnras/stae2143

\bibitem{Planck2020}
Planck Collaboration et al.  
Planck 2018 results. VI. Cosmological parameters.  
\textit{\aap} \textbf{641}, A6 (2020).
doi:10.1051/0004-6361/201833880

\bibitem{wollenberg2020formation}
Wollenberg, K. M. J., Glover, S. C. O., Clark, P. C. \& Klessen, R. S.  
Formation sites of Population III star formation: effects of rotation and turbulence on fragmentation.  
\textit{\mnras} \textbf{494}, 1871–1893 (2020).
doi:10.1093/mnras/staa289

\bibitem{tress2020simulations}
Tress, R. G. et al.  
Simulations of the Milky Way’s central molecular zone – I. Gas dynamics.  
\textit{\mnras} \textbf{499}, 4455–4478 (2020).
doi:10.1093/mnras/staa3120

\bibitem{krumholz2004embedding}
Krumholz, M. R., McKee, C. F. \& Klein, R. I.  
Embedding Lagrangian sink particles in Eulerian grids.  
\textit{\apj} \textbf{611}, 399–412 (2004).
doi:10.1086/421935

\bibitem{Enzo2014}
Bryan, G. L. et al.  
ENZO: An adaptive mesh refinement code for astrophysics.  
\textit{\apjs} \textbf{211}, 19 (2014).
doi:10.1088/0067-0049/211/2/19

\bibitem{Enzo2019}
Brummel-Smith, C. et al.  
ENZO: An adaptive mesh refinement code for astrophysics (Version 2.6).  
\textit{\joss} \textbf{4}, 1636 (2019).
doi:10.21105/joss.01636
    
\bibitem{regan2018rise}
Regan, J. A. \& Downes, T. P.  
Rise of the first supermassive stars.  
\textit{\mnras} \textbf{478}, 5037–5049 (2018).
doi:10.1093/mnras/sty1289

\bibitem{truelove1997jeans}
Truelove, J., K., Klein, R., I., McKee, C., F., Holliman, J., H., II, Howell, L., H., \& Greenough, J. A. 
The jeans condition: a new constraint on spatial resolution in simulations of isothermal self-gravitational hydrodynamics. 
\textit{\apj} \textbf{489}, L179 (1997).
doi:10.1086/310975

\bibitem{maeder2008physics}
Maeder, A.  
\textit{Physics, Formation and Evolution of Rotating Stars}.  
(Springer, 2009).

\bibitem{gatto2015modelling}
Gatto, A., Walch, S., Mac Low, M.-M., Naab, T., Girichidis, P., Glover, S. C. O., Wunsch, R., Klessen, R. S., Clark, P. C., Baczynski, C., Peters, T., Ostriker, J. P., Ibanez-Mejia, J. C., \& Haid, S.
Modelling the supernova-driven ISM in different environments.
\textit{\mnras} \textbf{449}, 1057–1075 (2015).
doi:10.1093/mnras/stv324

\bibitem{magg2022metal}
Magg, M., Schauer, A. T. P., Klessen, R. S., Glover, S. C. O., Tress, R. G., \& Jaura, O.
Metal Mixing in Minihalos: The Descendants of Pair-instability Supernovae.
\textit{\apj} \textbf{929}, 119 (2022).
doi:10.3847/1538-4357/ac5aac

\bibitem{smith2019cosmological}
Smith, M. C., Sijacki, D., \& Shen, S.
Cosmological simulations of dwarfs: the need for ISM physics beyond SN feedback alone.
\textit{\mnras} \textbf{485}, 3317–3333 (2019).
doi:10.1093/mnras/stz599

\bibitem{bondi1952spherically}
Bondi, H.
On spherically symmetrical accretion.
\textit{\mnras} \textbf{112}, 195 (1952).
doi:10.1093/mnras/112.2.195

\bibitem{dimatteo2005energy}
Di Matteo, T., Springel, V., \& Hernquist, L.
Energy input from quasars regulates the growth and activity of black holes and their host galaxies.
\textit{\nat} \textbf{433}, 604–607 (2005).
doi:10.1038/nature03335

\bibitem{dimatteo2008direct}
Di Matteo, T., Colberg, J., Springel, V., Hernquist, L., \& Sijacki, D.
Direct Cosmological Simulations of the Growth of Black Holes and Galaxies.
\textit{\apj} \textbf{676}, 33–53 (2008).
doi:10.1086/524921

\bibitem{sijacki2007unified}
Sijacki, D., Springel, V., Di Matteo, T., \& Hernquist, L.
A unified model for AGN feedback in cosmological simulations of structure formation.
\textit{\mnras} \textbf{380}, 877–900 (2007).
doi:10.1111/j.1365-2966.2007.12153.x

\bibitem{springel2005modelling}
Springel, V., Di Matteo, T., \& Hernquist, L.
Modelling feedback from stars and black holes in galaxy mergers.
\textit{\mnras} \textbf{361}, 776–794 (2005).
doi:10.1111/j.1365-2966.2005.09238.x

\bibitem{abramowicz2013foundations}
Abramowicz, M. A., \& Fragile, P. C.
Foundations of Black Hole Accretion Disk Theory.
\textit{\lrr} \textbf{16}, 1 (2013).
doi:10.12942/lrr-2013-1

\bibitem{sadowski2016energy}
Sadowski, A., Lasota, J.-P., Abramowicz, M. A., \& Narayan, R.
Energy flows in thick accretion discs and their consequences for black hole feedback.
\textit{\mnras} \textbf{456}, 3915–3928 (2016).
doi:10.1093/mnras/stv2854

\bibitem{Madau_2014}
Madau, P., Haardt, F., \& Dotti, M.
Super-critical Growth of Massive Black Holes from Stellar-mass Seeds.
\textit{\apjl} \textbf{784}, L38 (2014).
doi:10.1088/2041-8205/784/2/L38

\bibitem{dalla2012simulating}
Dalla Vecchia, C., \& Schaye, J.
Simulating galactic outflows with thermal supernova feedback.
\textit{\mnras} \textbf{426}, 140–158 (2012).
doi:10.1111/j.1365-2966.2012.21704.x

\bibitem{prole2022fragmentation}
Prole, L. R., Clark, P. C., Klessen, R. S., \& Glover, S. C. O.
Fragmentation-induced starvation in Population III star formation: a resolution study.
\textit{\mnras} \textbf{510}, 4019–4030 (2022).
doi:10.1093/mnras/stab3697

\bibitem{hartwig2015improved}
Hartwig, T., Glover, S. C. O., Klessen, R. S., Latif, M. A., \& Volonteri, M.
How an improved implementation of H\textsubscript{2} self-shielding influences the formation of massive stars and black holes.
\textit{\mnras} \textbf{452}, 1233–1244 (2015).
doi:10.1093/mnras/stv1368

\bibitem{clark2011formation}
Clark, P. C., Glover, S. C. O., Smith, R. J., Greif, T. H., Klessen, R. S., \& Bromm, V.
The Formation and Fragmentation of Disks Around Primordial Protostars.
\textit{\sci} \textbf{331}, 1040 (2011).
doi:10.1126/science.1198027

\bibitem{schauer2017formation}
Schauer, A. T. P., Regan, J., Glover, S. C. O., \& Klessen, R. S.
The formation of direct collapse black holes under the influence of streaming velocities.
\textit{\mnras} \textbf{471}, 4878–4884 (2017).
doi:10.1093/mnras/stx1915

\bibitem{Krumholz_2006}
Krumholz, M. R., McKee, C. F. \& Klein, R. I.
Bondi–Hoyle accretion in a turbulent medium.
\textit{\apj} \textbf{638}, 369–381 (2006).
doi:10.1086/498844

\bibitem{codezenodo}
Mehta, D. H., Regan, J. A., Prole. L.
Analysis Pipeline for Mehta et al. 202X
\textit{Zenodo} (2025)
doi:10.5281/zenodo.17894541

\bibitem{datafigshare}
Mehta, D. H., Regan, J. A., Prole. L.
Analysis datasets for Mehta et al. 202X
\textit{figshare} (2025)
doi:10.6084/m9.figshare.30857603


\end{thebibliography}

\end{document}